\documentclass[twocolumn, tighten, numberedappendix]{aastex631}
\pdfoutput=1 

\usepackage{amsmath,amstext}
\usepackage[T1]{fontenc}
\usepackage{apjfonts} 
\usepackage{graphicx}
\usepackage{xcolor}
\usepackage{color}
\usepackage{appendix}
\usepackage{microtype}
\usepackage{comment}
\usepackage{enumitem}
\usepackage{amssymb}

\newcommand{\WMAP}{\textsl{WMAP}}

\newcommand{\wmap}{{\WMAP}}
\newcommand{\Planck}{{\textsl{Planck}}}
\newcommand{\planck}{{\textsl{Planck}}}

\renewcommand{\ell}{\ensuremath{l}}
\newcommand{\be}{\begin{equation}}
\newcommand{\ee}{\end{equation}}

\newcommand{\beq}{\begin{equation}}
\newcommand{\eeq}{\end{equation}}
\newcommand{\beqa}{\begin{eqnarray}}
\newcommand{\eeqa}{\end{eqnarray}}

\def\ba{\begin{eqnarray}}
\def\ea{\end{eqnarray}}

\newcommand{\barr}{\begin{array}}
\newcommand{\earr}{\end{array}}

\renewcommand{\S}{Section~}

\usepackage[flushleft]{threeparttable}
\usepackage{array}

\usepackage{savesym}
\savesymbol{tablenum}
\usepackage[separate-uncertainty=true, multi-part-units=single]{siunitx}
\restoresymbol{SIX}{tablenum}

\usepackage{gensymb}
\usepackage{float}
\usepackage{makecell}
\usepackage{hyperref}
\usepackage[hyphens]{url}

\providecommand{\sorthelp}[1]{}

\usepackage{natbib}

\shorttitle{}

\begin{document}

\title{Polarized Synchrotron Foreground Assessment for CMB Experiments}

\correspondingauthor{J. L. Weiland}
\email{jweilan2.jhu.edu}

\author[0000-0003-3017-3474]{Janet L. Weiland}   
\affiliation{The William H. Miller III Department of Physics \& Astronomy \\
Johns Hopkins University \\
3400 North Charles Street \\
Baltimore, MD 21218, USA}

\author[0000-0002-2147-2248]{Graeme E. Addison}
\affiliation{The William H. Miller III Department of Physics \& Astronomy \\
Johns Hopkins University \\
3400 North Charles Street \\
Baltimore, MD 21218, USA}

\author[0000-0001-8839-7206]{Charles L. Bennett}
\affiliation{The William H. Miller III Department of Physics \& Astronomy \\
Johns Hopkins University \\
3400 North Charles Street \\
Baltimore, MD 21218, USA}

\author[0000-0002-1760-0868]{Mark Halpern}
\affiliation{Department of Physics and Astronomy \\
University of British Columbia \\ 
Vancouver, BC  Canada V6T 1Z1}

\author[0000-0002-4241-8320]{Gary Hinshaw}
\affiliation{Department of Physics and Astronomy \\
University of British Columbia \\ 
Vancouver, BC  Canada V6T 1Z1}

\begin{abstract}    

Polarized Galactic synchrotron emission is an undesirable foreground for cosmic microwave background (CMB) experiments
observing at frequencies $< 150$~GHz.  
We perform a combined analysis of observational data at 1.4, 2.3, 23, 30 and 33~GHz to quantify the spatial variation of
the polarized synchrotron spectral index, $\beta^{pol}$, on $\sim3.5^\circ$ scales.  We compare results from different data combinations to address limitations and inconsistencies present in these public data, and form a composite map of $\beta^{pol}$. 
Data quality masking leaves 44\% sky coverage (73\% for $|b|> 45^\circ$). 
Generally $-3.2 < \beta^{pol} \lesssim -3$ in the inner Galactic plane and spurs, but the Fan Region in the outer Galaxy has a flatter index.
We find a clear spectral index steepening with increasing latitude south of the Galactic plane with $\Delta \beta^{pol}=0.4$, and a smaller steepening of $0.25$ in the north. Near the south Galactic pole the polarized synchrotron spectral index is $\beta^{pol} \approx -3.4$. Longitudinal spectral index variations of $\Delta \beta^{pol} \sim 0.1$ about the latitudinal mean are also detected.
Within the BICEP2/Keck survey footprint, we find consistency with a constant value, $\beta^{pol} = -3.25 \pm 0.04$ (statistical) $\pm 0.02$ (systematic).
We compute a map of the frequency at which synchrotron and thermal dust emission contribute equally to the total polarized foreground.
The limitations and inconsistencies among datasets encountered in this work
make clear the value of additional independent surveys at multiple frequencies, especially between $10-20$~GHz,
provided these surveys have sufficient sensitivity and control of instrumental systematic errors.
\end{abstract}

\section{Introduction}

The cosmic microwave background (CMB) has been crucial to cosmology. All CMB experiments must ensure that foreground contamination is minimized and upper limits are well-estimated, as this often sets the ultimate limits on cosmological results. With the current and future observational focus on CMB polarization measurements, characterization of
polarized foregrounds is a key aspect of CMB studies.

Galactic polarized foregrounds are dominated by synchrotron emission at lower frequencies and thermal dust emission at higher frequencies with a cross-over at $\sim 80$ GHz, depending on sky location (e.g., \citealt{page/etal:2007, bennett/etal:2013, planck/10:2015}).
Other sources of low frequency polarized emission are theoretically possible, and must happen at some level, but have not yet been detected. This includes polarized spinning dust emission, which is usually considered to arise from radius $a\lesssim 1$ nm rapidly rotating dust grains with an electric dipole moment \citep{erickson:1957, draine/lazarian:1998,hensley/draine:2017}. While spinning dust emission has been detected, its polarization has not \citep{genova-santos/etal:2017, 
dickinson/etal:2018}. \cite{draine/hensley:2016} suggest that the spinning dust polarization will be negligible due to quantization of the vibrational energy levels in the emitting grains, which would exponentially suppresses their alignment. In this paper we attribute all of the low frequency polarized emission to the synchrotron mechanism.

Synchrotron emission is generated by a population of ultra-relativistic electrons that spiral around magnetic field lines in our Galaxy. For a power-law population distribution of electron energies $N(\gamma)\propto \gamma^{-p}$, where $\gamma$ is the relativistic Lorentz factor, the maximum degree of polarization 
(polarization fraction)
in a uniform magnetic field is $P_{\rm max}=(3p+3)/(3p+7)$ \citep{rybicki/lightman:1986}. The Galactic magnetic field is typically $\sim 6 \, \mu$G \citep{beck:2001}
and for $p=3.0$ the spectral index is $\beta=-(p+3)/2 = -3.0$ and the maximum degree of polarization $P_{\rm max}$ is 75\%. However, this high degree of polarization is almost never observed for many reasons, including the pitch angle distribution of the emitting electrons along a line of sight that effectively depolarizes the observed emission. 
In units of antenna temperature, the spectral index $\beta$ is defined by $T\propto \nu^{\beta}$.  
We use the notation $\beta^{pol}$ in subsequent sections to emphasize that the spectral index specifically refers to the polarized signal. 
The spectral index may vary both in the frequency and spatial domains.  Mechanisms such as synchrotron self-absorption, along
with cosmic ray electron propagation and energy loss, are contributing factors \citep{strong/orlando/jaffe:2011}.

Single region
foreground template removal methods have been adequate for cosmological analyses of large sky areas
surveyed by previous experiments such as 
\WMAP\footnote{Wilkinson Microwave Anisotropy Probe} and \Planck\ 
(e.g., \citealt{hinshaw/etal:2013, planck/06:2018}).
However, \citet{errard/stompor:2019} 
noted that spurious detections of the tensor-to-scalar ratio $r$ of order $0.01$ could result from using a single set
of spectral parameters to clean foregrounds over the entire observed sky.
The demands on accurate foreground removal will only grow in the future with deeper measurements and more aggressive goals. 
\citet{osumi/etal:2021} recently showed that the polarized dust spectral index variations are not currently well enough known for ambitious new experiments. We now consider how well we can characterize the polarized synchrotron foreground emission with current public observational data.
This study is intended not only to assess the current status of our knowledge of this foreground, but also to highlight those areas where
additional observations would provide constraints that the current data do not.
We cite previous work of particular relevance to this analysis in the body of the paper.
Additionally, Table ~1 of \citet{jew/grumitt:2020} provides a summary of references to earlier determinations of the polarized synchrotron
spectral index.

This paper is organized as follows. We introduce the observational data we use in Section~\ref{sec:data} along with a discussion of Faraday rotation and depolarization. We present our data analysis of the Galactic plane region together with the associated spur regions in 
Section~\ref{sec:hisig}. These regions have relatively higher signal-to-noise ratios (SNR) but also higher Faraday rotation than other sky regions. In Section~\ref{sec:offplane} we present our data analysis of the higher Galactic latitude regions, which have lower SNR. We consider several types of data analyses using differing combinations of data. In Section~\ref{sec:composite} we create a composite polarized synchrotron spectral index map and use it to derive results, while also illustrating the substantial limitations of the current data for these purposes.
Difficulties we encountered in connecting polarization and low frequency intensity data are presented in Section~\ref{sec:intensity}.
Section~\ref{sec:implications} discusses implications of our findings for synchrotron foreground removal.
Finally, we summarize our conclusions in Section~\ref{sec:concl}.

\section{Observational Data} \label{sec:data}

\subsection{Polarized sky data}
At present there are only a few publicly available polarized maps of the sky at frequencies dominated by synchrotron emission.
The \WMAP\ \citep{bennett/etal:2013} and \planck\ \citep{planck/01:2018} space missions provide full sky coverage in the $23 - 44$ GHz range.  These offer the advantage of precise gain calibration with minimal Faraday rotation, counterbalanced by relatively low SNR in some high Galactic latitude directions. 
We limit our investigation to the three frequencies with the highest SNR: \wmap\ 23~GHz (K-band),
\planck\ Low Frequency Instrument (LFI) 30~GHz, and \wmap\ 33~GHz (Ka-band).  There are multiple versions of \planck\ LFI 30~GHz maps available.  These include the 2018 Public Release 3 (PR3, \citealt{planck/01:2018}), 
the 2020 Public Release 4 (PR4, aka NPIPE, \citealt{npipe:2020}), and the BeyondPlanck \citep{beyondplanck_pipeline:2020} maps.
We discuss our choices of which maps to use in the data analysis sections.

\begin{figure}[th]
    \centering
    \includegraphics[width=2.5in]{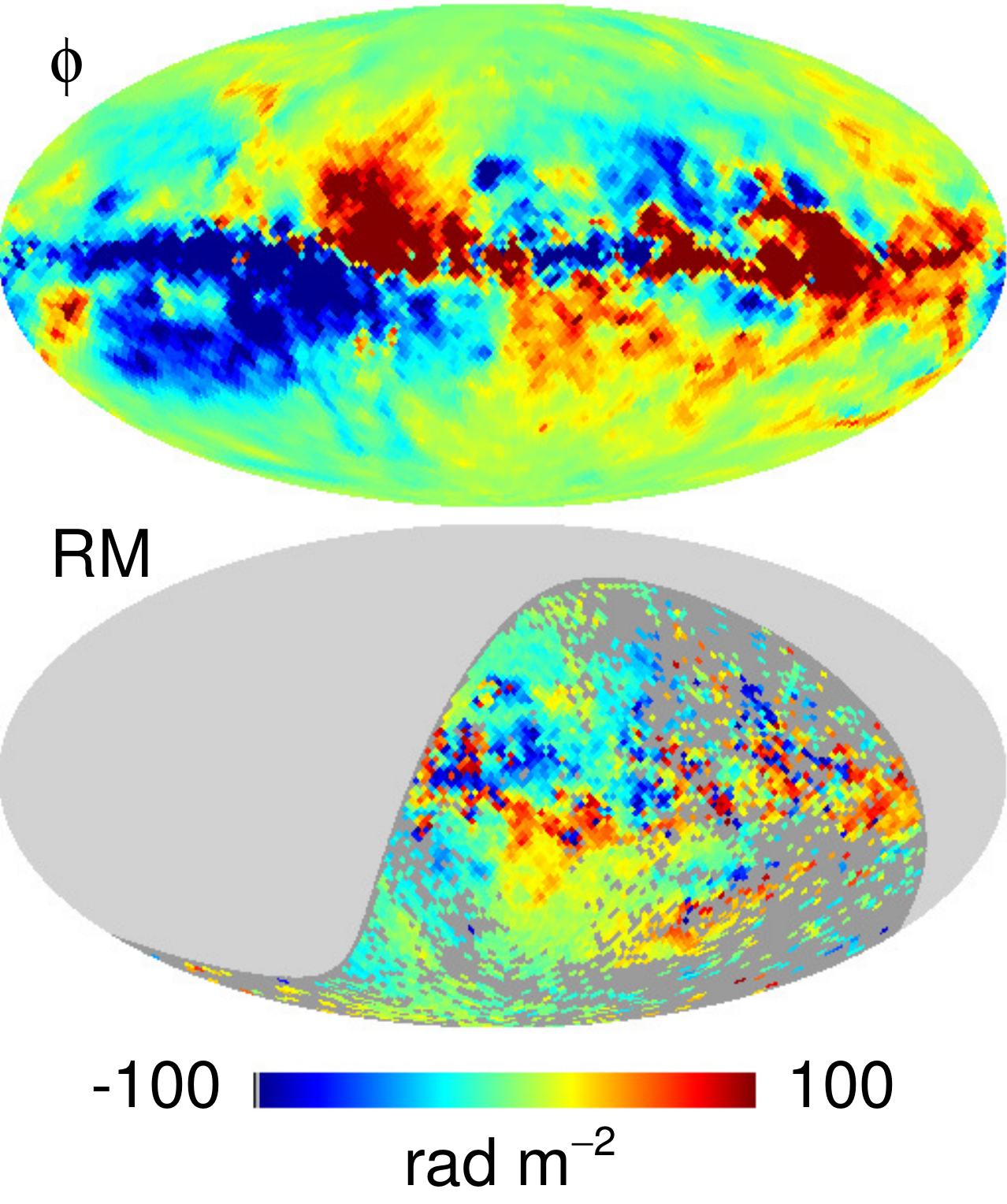}
    \caption{{\it{Top:}} The Faraday depth ($\phi$) map of \citet{hutschenreuter/etal:2021}. 
    {\it{Bottom:}} The RM map of \citet{carretti/etal:2019} determined from 2.3, 23 and 30 GHz polarization observations.  Light gray indicates areas not included in the S-PASS 2.3 GHz survey.  Dark grey indicates areas for which no RM was computed due to 
    inconsistency between \wmap\ and \planck\ polarization angles.
    For a given line of sight, $\phi$ and $RM$ tend to be in greater accord at higher latitudes and regions of lower $\phi$ magnitude.
    }
    \label{fig:rm_maps}
\end{figure}

Ground based observations at lower frequencies include S-PASS\footnote{S-band Polarisation All Sky Survey} (2.3 GHz, $\delta < -1^\circ$, \citealt{carretti/etal:2019}),
DRAO\footnote{Dominion Royal Astrophysical Observatory} (1.41 GHz, $\delta > -29^\circ$, \citealt{wolleben/etal:2006}), Villa Elisa (1.435 GHz, $\delta < -10^\circ$, \citealt{testori/etal:2008}).  The latter two 1.4 GHz surveys may be merged to form a full-sky map, but were not originally projected in HEALPix\footnote{Hierarchical Equal Area isoLatitude Pixelisation} \citep{gorski/etal:2005}).  We use the DRAO-only HEALPix projection of
\citet{laporta/etal:2005}, as well as the merged all-sky version from 
Centre d'Analyse de Donn\'{e}es Etendues (CADE)\footnote{\url{http://cade.irap.omp.eu/dokuwiki/doku.php}}.

Polarization angle conventions are not inherently consistent between all the above datasets. While the \wmap\ and \planck\ products follow the HEALPix convention, those of the S-PASS and 1.4 GHz surveys follow the IAU convention.  The sign of Stokes U changes when converting between the two conventions. 
Figures shown in this paper follow the HEALPix convention,
and unless otherwise noted, maps are shown as Mollweide
projections in Galactic coordinates, centered on $l=0^\circ, b=0^\circ$.
Temperature units are also not inherently consistent between
datasets.  Both \Planck\ and \wmap\ have adopted thermodynamic units, while the ground-based data are in antenna (Rayleigh Jeans)
temperature.  In general, we show each in their native units, but the conversion factors between the two systems are relatively
small (a few percent) for these frequencies.  

In comparison to the lower frequency datasets,
noise characteristics of \wmap\ and \planck\ maps are somewhat complex. \WMAP\ noise
includes pixel-pixel covariance at lower resolutions as well as
some large-scale modes with enhanced uncertainty
\citep{jarosik/etal:2003, jarosik/etal:2007, jarosik/etal:2011}. \planck\ noise includes non-Gaussian instrumental effects and systematic residuals \citep{planck/02:2018}.
We therefore also make use of ancillary products such as data splits, covariance matrices and simulations
provided by the instrument teams for noise characterization, as described in individual analysis sections.

\vspace{7mm}
\subsection{Faraday rotation measures}\label{sec:rm_data}

Both the 1.4 and 2.3~GHz data are significantly affected by Faraday rotation and depolarization.
The rotation measure ($RM$) defines a plane of polarization rotation angle $\theta$ per
line-of-sight at a given frequency, assuming a wavelength-squared dependence: $\theta = RM~\lambda^2 $
\citep{burn:1966}.
A rotation matrix with $2\theta$ as the argument, applied to the ``true'' signal as originally emitted, describes
the observed Faraday-rotated Stokes Q and U signals after passage through the ISM
\citep{vinyajkin:2004, fuskeland/etal:2019, oppermann/etal:2015}:

\begin{equation}
    \begin{bmatrix}
    Q \\
    U
    \end{bmatrix}
    ^{\rm{obs}}
    =
    \begin{bmatrix}
     \cos{2\theta} & 
     \sin{2\theta} \\
     -\sin{2\theta} & 
     \cos{2\theta}
    \end{bmatrix}
    \begin{bmatrix}
     Q \\
     U
    \end{bmatrix}
    ^{\rm{true}}.
\end{equation}

The polarized intensity $P = \sqrt{Q^2 + U^2}$ is invariant under pure rotation.  
However, construction of P from Q and U maps with a significant noise component introduces a positive noise bias
which must be accounted for.  Propagation of uncertainties is more straightforward when using Q and U directly,
with the necessary accompanying Faraday rotation corrections.  
In our analyses of Q and U maps, we make use of the all-sky Faraday depth ($\phi$) map of \citet{hutschenreuter/etal:2021} determined from observations of
extragalactic sources, and the $RM$ map of \citet{carretti/etal:2019} determined from S-PASS, \wmap\ and \planck\ data for specific lines-of-sight.  These two maps are shown in the top and bottom panels of
Figure~\ref{fig:rm_maps} respectively.
Under thin-screen conditions, $RM$ and $\phi$ are equivalent.  
However, in the case of cumulative multiple rotations along the propagation path, depolarization is likely to occur with
$RM \neq \phi$ \citep{wolleben2/etal:2010, hutschenreuter/etal:2021}.  

Faraday rotation and depolarization can greatly reduce the amount of detectable polarized emission, lowering the
observed polarization fraction and producing a shallower synchrotron spectral index than would be observed in its absence.
Throughout this paper, we either mask to avoid regions subject to depolarization, apply RM corrections under conditions where
$\phi \approx RM$, or use frequencies for which Faraday effects are not significant.

\section{Analysis of Galactic Plane and Spur Regions}\label{sec:hisig}

The highest SNR regions available for determinations of $\beta^{pol}$ are located close
to the Galactic plane and in the extensions (spurs) reaching north and south off the plane near the Galactic center. 
However, Faraday depolarization near the Galactic plane limits $\beta^{pol}$ analyses in these regions to frequencies
$> 5-10$ GHz, with \Planck\ and \WMAP\ data providing full sky coverage and sub-percent absolute
calibration. Some previous results using \planck\ LFI or \wmap\ data and specifically including plane and spur regions are 
\citealt{fuskeland/etal:2014, vidal/etal:2015, jew/grumitt:2020} and \citet{beyondplanck/fg:2020}.
When combined with \wmap\ K-band, QUIET\footnote{QU Imaging ExperimenT} 43~GHz observations of planar regions designated G-1 ($l = 329^\circ$) and G-2 ($l = 0^\circ$) result in $\beta^{pol} \sim -2.9$ on-plane ($\vert b\vert \leq 2.5^\circ$) and $\sim -3.07$ to $-3.14$ off-plane ($2.5^\circ < \vert b\vert <10^\circ$) \citep{quiet:2015}.

We revisit a determination of $\beta^{pol}$ in these high SNR regions, using pairs of frequencies: (i) the 9-year \wmap\ 
23 ~GHz (K-band) and 33~GHz (Ka-band) maps, and separately (ii) the \wmap\ 23~GHz and \planck\ public release 3 (PR3) LFI 30 GHz maps.  Ideally, one would wish to perform
a combined three-frequency fit to optimally reduce instrumental noise contributions, but for reasons we discuss below, this is not a straightforward exercise.
  
We use a linear correlative analysis method similar to that described by \citet{fuskeland/etal:2014} for each data frequency pair.  Stokes Q and U maps at each frequency are first smoothed to $2^\circ$ FWHM, and degraded to HEALPix $N_{side}=32$ pixels, or about $1.8^\circ$ on a side. We then fit a line of zero intercept through the combined Q and U
data points within a $\sim 7^\circ$ superpixel (HEALPix $N_{side}=8$).  Converted to antenna temperature units, the slope of this line, $m$, determines the spectral index: $\beta^{pol} = {\rm{log}}(m)/{\rm{log(\nu_y/}}\nu_x)$, where  $\nu_x$ and $\nu_y$ are the frequencies associated with the data plotted on the two axes. 
There are a total of 32 data pairs possible per superpixel (16 Q, 16 U).  
All pairs are used in each superpixel fit except for a small number of cases in which bright supernova remnants close to the plane were excluded
(Tycho, 3C58, W51, W63, Pup A, Cas A, Tau A and MSH15\_56).
Since the instrumental noise uncertainties are comparable for both frequencies,
statistical weighting along both
axes points is applied.  The uncertainty of the fit slope parameter $\sigma_m$ is used to compute the statistical uncertainty in $\beta^{pol}$:
${\sigma(\beta^{pol}) = \sigma_m/[m\; {\rm{log}}(\nu_y/\nu_x)]}$
as in e.g.  equation 11 of \citet{fuskeland/etal:2014}.

Data uncertainties used in the linear fits are derived from the diagonal QQ and UU pixel variances appropriate to each map.  However, these per-pixel uncertainties do not fully encompass
the more complex noise characteristics of \wmap\ and \planck\ data (see Section~\ref{sec:data}).  
We use simulations to verify that the use of the diagonal uncertainties in the fit does not bias  recovery of the true
parameters and their uncertainties.  Simulated maps include realistic sky signals and 
independently generated instrument noise that more completely characterizes the full noise properties
of the data.

The simulation suite consists of Q and U maps produced from the sum of synchrotron, CMB and instrument noise components.
We use the synchrotron sky model from the \planck\ Full Focal Plane (FFP10) simulations, adopting a 23~GHz Q and U signal level and then use a spatially constant $\beta^{pol}$ of either $-3.15$ or $-2.8$ to extrapolate to the
other frequencies.  Two choices of $\beta^{pol}$ were used to verify that results are not sensitive to the chosen spectral index
value.
The synchrotron component does not vary with realization, but the CMB and instrument noise components are different for each realization.  The CMB component adopts a standard $\Lambda$CDM cosmology from the \planck\ PR3 public chains\footnote{chain identifier: base\_plikHM\_TTTEEE\_lowl\_lowE\_lensing}.  
The inclusion of the CMB component is not essential, since its contribution is
completely subdominant to the foreground signal and noise.
\WMAP\ 23 and 33~GHz noise maps are computed from bootstrapped 
samples taken from the population of single-year null difference maps, scaled to 9-year noise levels.
The \planck\ 30 GHz noise is taken from the FFP10 simulations.  We apply the sample smoothing and
down-sampling to the simulations as with the data.
Uncertainties from the simulations generally agree with those derived from the data within $\sim 5$\%.

\begin{figure}[t]
\centering
\includegraphics[width=2.5in]{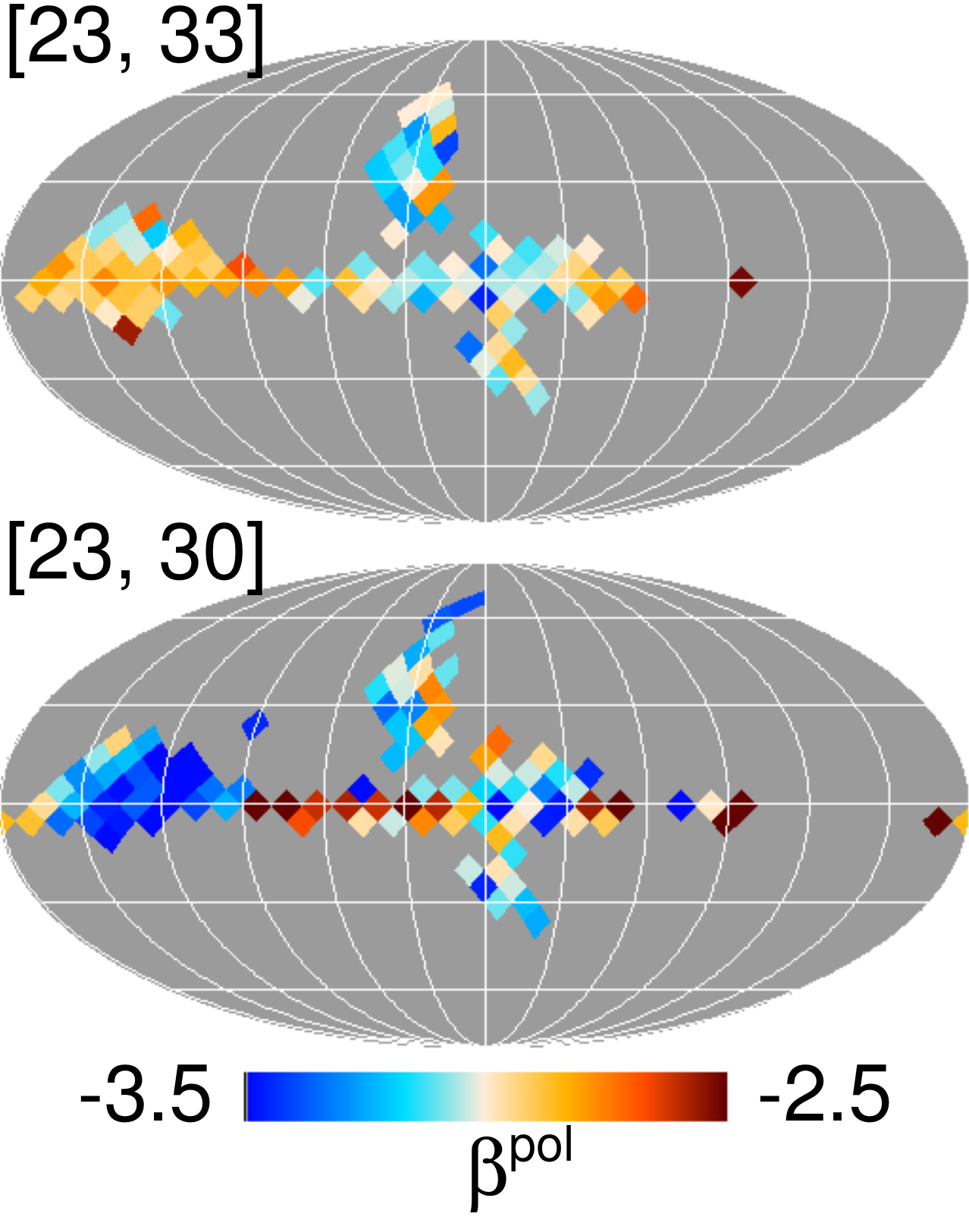}
\caption{{\it{Top:}} Polarized spectral index $\beta^{pol}$ computed from \wmap\ 23~GHz and 33~GHz 9-yr Q and U maps, in Galactic coordinates. 
White coordinate grid lines are in $30^\circ$ increments, with the convention 
${l,b=(0^\circ,0^\circ)}$ is at the center, ${(90^\circ,0^\circ)}$ mid-way to the left, and 
${(270^\circ,0^\circ)}$ mid-way to the right.
{\it{Bottom:}} The $\beta^{pol}$ map computed from \wmap\ 23~GHz and
\planck\ LFI 30 GHz from Public Release 3 (PR3).  Pixel sizes correspond to HEALPix $N_{side}=8$ ($\sim 7^\circ$ on a side). Pixels for which $\sigma (\beta^{pol}) > 0.2$ are shown as the gray masked region.  There are clear visual differences between the two determinations: those in the Fan Region (on the left at $l \sim 140^\circ$, 
where the colors shift from mostly orange to mostly dark blue) are statistically significant, but
are attributable to large-scale systematic differences between \wmap\ and \Planck\ LFI maps. 
As discussed by \citet{planck/02:2018}, the large-scale systematic modes are primarily attributable to the 30~GHz map.}
\label{fig:beta_kka_k30}
\end{figure}

Results of the $\beta^{pol}$ determination using the linear correlation method for \WMAP\ 23 and 33~GHz bands are shown in the top panel of Figure~\ref{fig:beta_kka_k30}.  
Only pixels with an uncertainty $\sigma(\beta^{pol}) \leq 0.2$ are shown; all others are excluded and shown in
gray.  We chose this limit for reliable determinations based on our simulations.  A consequence is that
at this pixel resolution, determinations are not included for fainter portions of the outer Galactic plane.
Uncertainties are typically lowest close to the Galactic plane where the SNR is highest, and those in the northern
and southern spurs are more typically near $0.15-0.2$.  The top panel shows $\beta^{pol} \lesssim -3$ over much
of the analyzed area, with a somewhat flatter index in the Fan Region (an extended bright region in the Galactic plane centered near $l \sim 140^\circ$; see e.g. \citealt{hill_fan/etal:2017}, \citealt{west/etal:2021} for possible origins).

\begin{figure}[ht]
\centering
\includegraphics[width=2.5in]{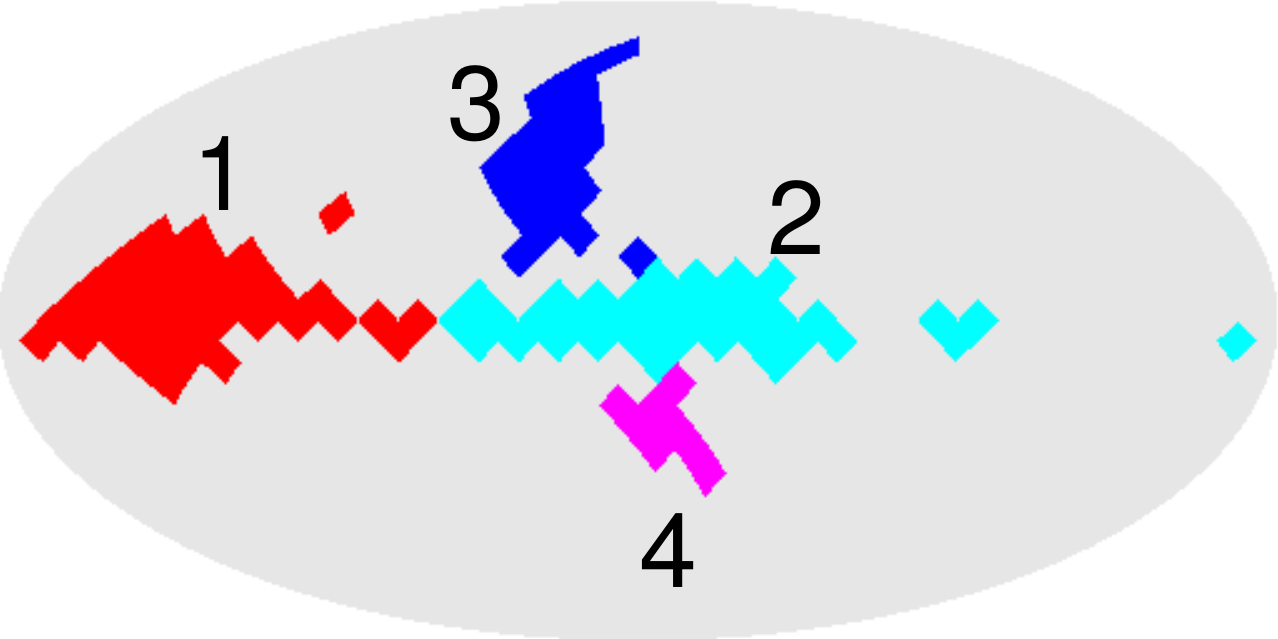}
\includegraphics[width=3.5in, trim={0.75cm 0 0 0}, clip]{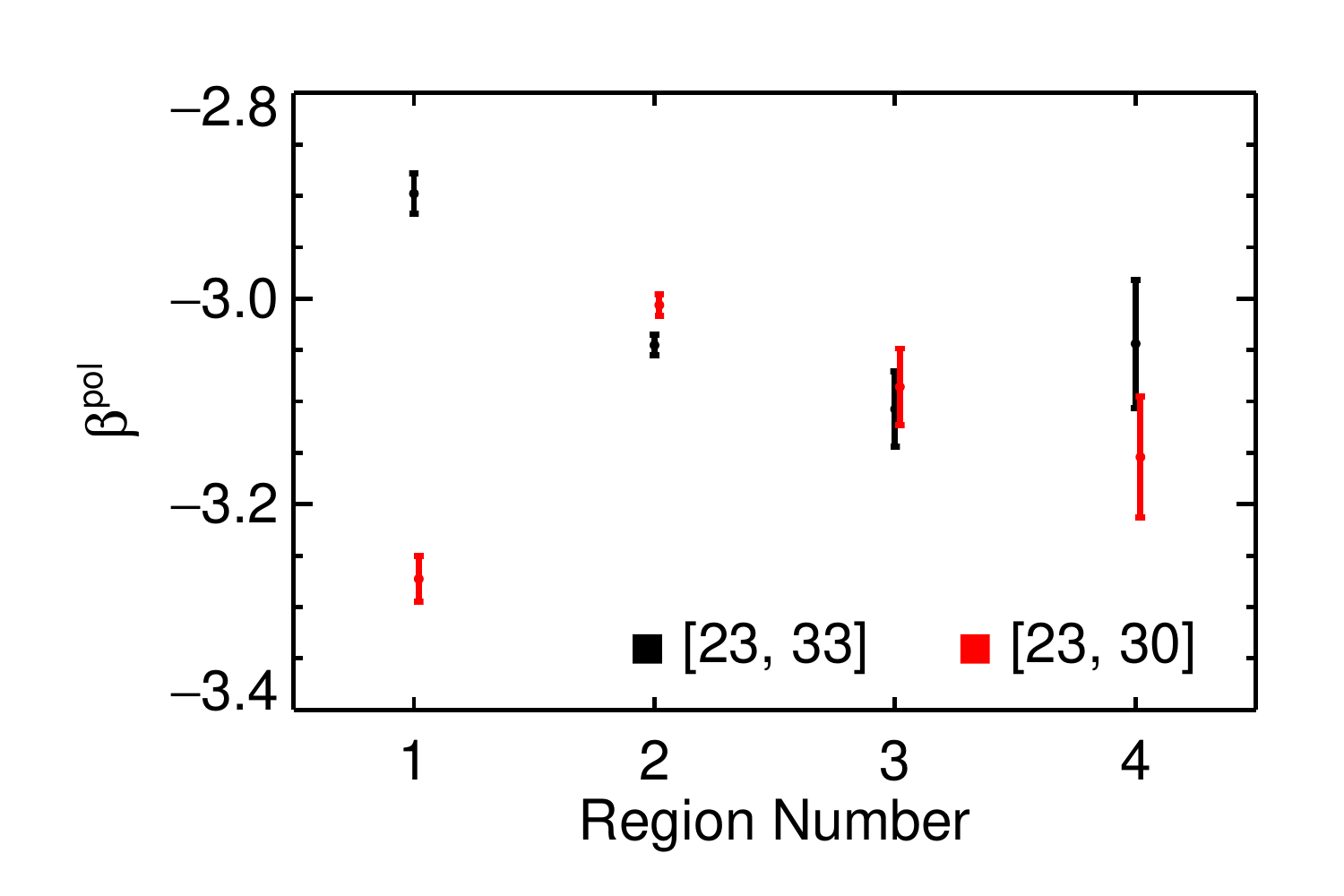}
\caption{{\it{Top:}} We group the analyzed pixels in Figure~{\ref{fig:beta_kka_k30}} into four larger regions,
with the Fan Region encompassed in region~1 (red), the inner Galaxy in region~2 (cyan), the northern spur in region~3 (blue), and the southern extension in region~4 (magenta).
{\it{Bottom:}} For each region, the weighted average of $\beta^{pol}$ and its statistical uncertainty (Table~\ref{tab:region_tab})
are computed separately for the [23,33] and [23,30] combinations and shown in black and red, respectively. Red and black points are
slightly offset from each other along the x-axis for visual clarity. Mean $\beta^{pol}$ values in
regions 2, 3 and 4 are roughly consistent between the two combinations, while those for region 1 are significantly different. 
}
\label{fig:beta_regions}
\end{figure}

\begin{figure}
    \centering
    \includegraphics[width=2.5in]{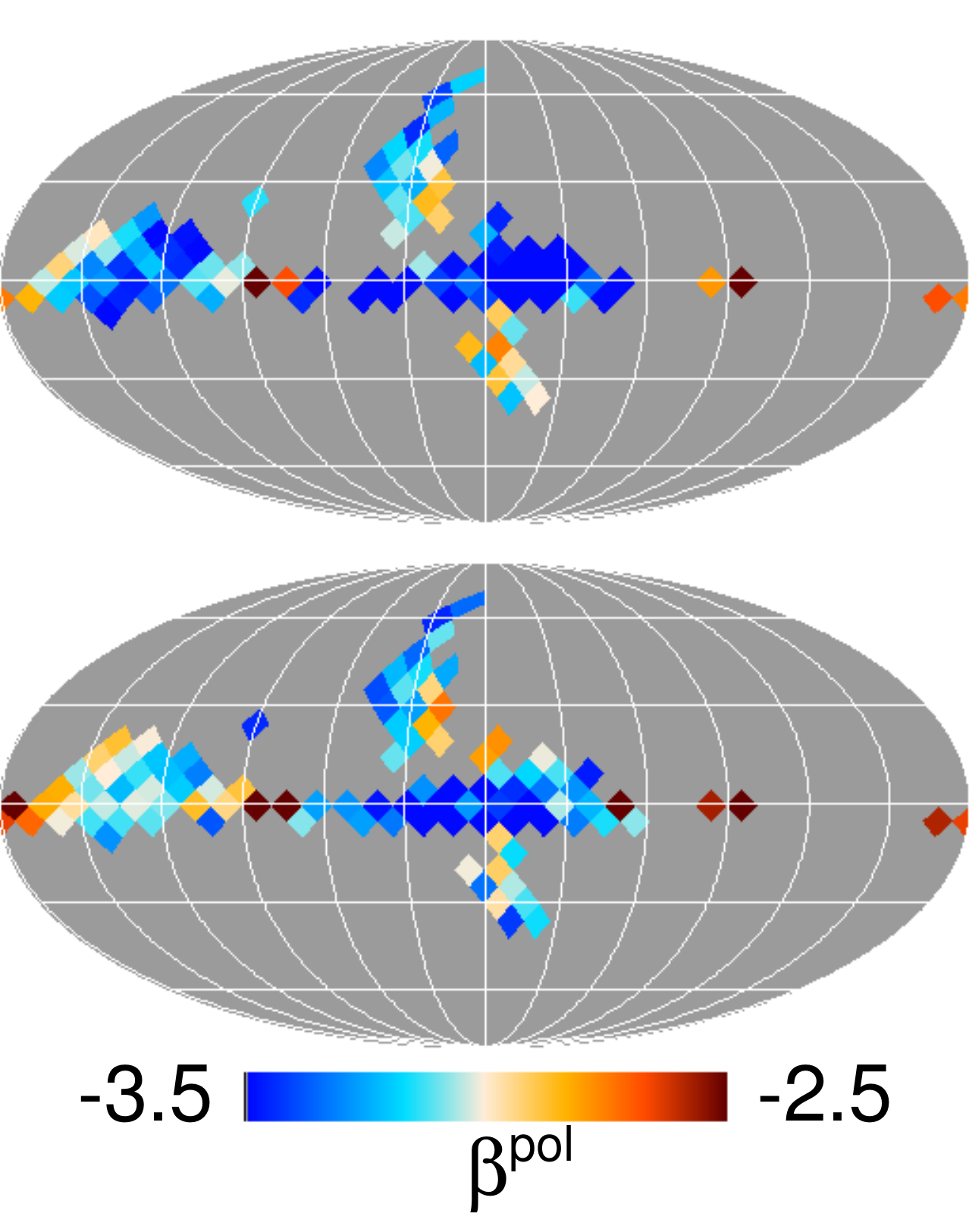}
    \caption{Maps of $\beta^{pol}$ derived from [23,30] in which the 23~GHz data is unchanged, but
    the 30~GHz band is either from the PR4 (NPIPE) release (top panel)
    or the BeyondPlanck release (bottom panel). Differences between the three [23,30] variations
    (two shown here and one 
    in Figure~\ref{fig:beta_kka_k30} using PR3 30~GHz data) are solely the result of different algorithmic
    choices in processing of the 30~GHz band.  Regions near the plane are the most affected.
    }
    \label{fig:k30_np_bp}
\end{figure}

In the bottom panel of Figure~\ref{fig:beta_kka_k30}, the $\beta^{pol}$ map derived using the same methodology
is shown for the \wmap\ 23 and LFI 30~GHz frequency pair.  There are similarities in the map with the [23,33] result,
but also clear differences.  The two most visually striking differences are within the plane: a steeper
[23,30] spectral index near the Fan Region, and a string of shallower index pixels running across the plane
near the inner Galaxy.  
We quantify the significance of these differences through the use of four larger regions as defined
in the top of Figure~\ref{fig:beta_regions}.  These regions correspond to: (1) an outer plane region dominated by the Fan Region,
(2) an inner plane dominated region, and northern (3) and southern (4) extensions off the plane.  For both the [23,30]
and [23, 33] analyses, weighted means and associated uncertainties for each region are given in Table~\ref{tab:region_tab}.  The results in Table~\ref{tab:region_tab} are plotted in the bottom panel of Figure~\ref{fig:beta_regions}.  The figure illustrates high statistical significance of the difference in 
$\beta^{pol}$ between the [23,33] and [23,30] determinations for region 1.  

There is no likely physical mechanism that suggests that the region 1 results for [23,30] and [23,33] are
both correct.
Faraday depolarization is not a significant
factor at these frequencies, and furthermore would act to produce a shallower [23,30] index compared to
[23,33] rather than the steeper index shown in Figure~\ref{fig:beta_regions}.  
The main beam FWHM of the 30~GHz and 33~GHz radiometers are similar (33$'$ and 40$'$ respectively), minimizing resolution-induced
changes in the sampled physical conditions.  Beam depolarization and spectral decoherence differences, rooted in magnetic field and synchrotron
SED variations across the beam, are minimized given that the 30 and 33~GHz 
bandpasses overlap
between $29-31$~GHz, and the frequency lever-arm between bands is not large. 
Abrupt changes in the electron energy spectrum are also unlikely at these frequencies
\citep{strong/orlando/jaffe:2011, orlando/strong:2013}.

A non-physical origin for the Region 1 discrepancy is more compelling.
Large-scale systematic differences between \WMAP\ and \Planck\ LFI polarization maps have been
noted in previous publications, e.g., \citet{weiland/etal:2018, planck/10:2015}.
\citet{planck/02:2018} show that these large-scale systematic differences 
arise primarily from  
an unconverged iterative (gain calibration + sky model) solution in the PR3 LFI processing pipeline.  
That portion 
of the large-scale gain uncertainty modes induced in the LFI polarization maps that intersects with region 1
directly correlates with the area of highest $\beta^{pol}$ discrepancy, and is of correct sign to produce
the steeper [23,30] $\beta^{pol}$.

\begin{table}[ht]
\centering
\caption{$\beta^{pol}$ for Plane and Spur Regions}
\label{tab:region_tab}

\begin{tabular}{c c c}
\hline
Region Number\footnote{see Figure~\ref{fig:beta_regions} for region definition.} & [23, 33]  &  [23, 30] \\ [0.5ex]
\hline\hline
   1  &  $-2.90 \pm 0.020$ & $-3.27 \pm 0.022$ \\
   2  &  $-3.05 \pm 0.010$ & $-3.01 \pm 0.011$ \\
   3  &  $-3.11 \pm 0.036$ & $-3.09 \pm 0.037$ \\
   4  &  $-3.04 \pm 0.062$ & $-3.15 \pm 0.059$ \\
 \hline  
\end{tabular}

\end{table}

Although we have primarily discussed results from either [23,33] or \planck\ PR3 data in combination with 23~GHz, 
we also performed some exploratory
computations in which either the NPIPE or BeyondPlanck 30 GHz maps were substituted for
the PR3 30 GHz maps in the [23,30] combination. 
We show these results in Figure~\ref{fig:k30_np_bp}.
The higher latitude regions 3 and 4 have mean values 
near $-3$ to $-3.1$, i.e., consistent with those already
shown in Table~\ref{tab:region_tab} and Figure~\ref{fig:beta_regions}.
However, results within the plane regions 1 and 2 are variable and depend solely on the version of 30~GHz maps
used in the computation, since the same 23~GHz data are used in each computation.
Using either one of these alternate LFI processings indicate substantially steeper $\beta^{pol}$ values in the inner plane region~2, with means close to  $-3.4$ or $-3.5$.  
Arguments against a physical interpretation, presented earlier for region 1, also apply to region 2.
Furthermore, the independent [23,43] $\beta^{pol}$ determination of \citet{quiet:2015} using QUIET data (see beginning of this section)
support a spectral index between -2.9 and -3.1 for this region, in agreement with the [23,33] result.
With the continued presence of large-scale modes of similar morphology to those in PR3, along with reliance on a sky model for certain aspects 
of LFI processing (e.g., temperature to polarization leakage) and complex nature of
emission within the Galactic plane,
we suspect that $\beta^{pol} \lesssim -3.3$ in region 2 does not reflect true sky variations but rather data processing
inconsistencies. 
Given that the polarized spectral index results depend significantly on processing differences of the same 30 GHz data set, we favor the use of the more stable and reliable 33 GHz data.
We 
adopt the $\beta^{pol}$ determined from the [23,33] frequency
pair as our best estimate for these high SNR regions.

As a function of longitude within the Galactic plane, the [23,33] regional map shows a trend toward a slightly steeper spectral index in the inner Galaxy compared to
the Fan Region, at a signficance of $\sim 3\sigma$.
A similar trend has been noted in previous analyses using different regional segmentations of the plane, e.g. \citet{fuskeland/etal:2014} and \citet{beyondplanck_pipeline:2020}.  
These trends address large scales and do not represent the full complexity of the Galactic plane.
The spatial resolution of the $\beta^{pol}$ map has been constrained primarily by SNR in the
\WMAP\ 23 and 33~GHz bands.  While the 23~GHz beam would allow a common resolution of $\sim1^\circ$,
the main effect of using $\sim 1^\circ$ pixels on this paper's results would be one of increased uncertainty.  Variations in electron energy spectral index and polarization fraction across a pixel that 
could be detectable at this higher resolution (e.g., \citealt{padovani/etal:2021}) require additional data for a meaningful result.

In terms of latitudinal dependence, 
there is no clear trend based solely on the results for regions 3 and 4. 
Emission from regions 3 and 4 are likely dominated by local structures (e.g., the North Polar Spur \citealt{vidal/etal:2015, west/etal:2021}).
Mean $\beta^{pol}$ values in these regions are consistent with mean high latitude values derived 
over very large sky fractions using \wmap\ and/or \planck\ data.
We more completely
address the higher latitude data in Section~\ref{sec:offplane}.

\section{Analysis of lower SNR off-plane regions} \label{sec:offplane}

\begin{figure*}
    \centering
    \includegraphics[width=6in,trim={0 0 0 1cm}, clip]{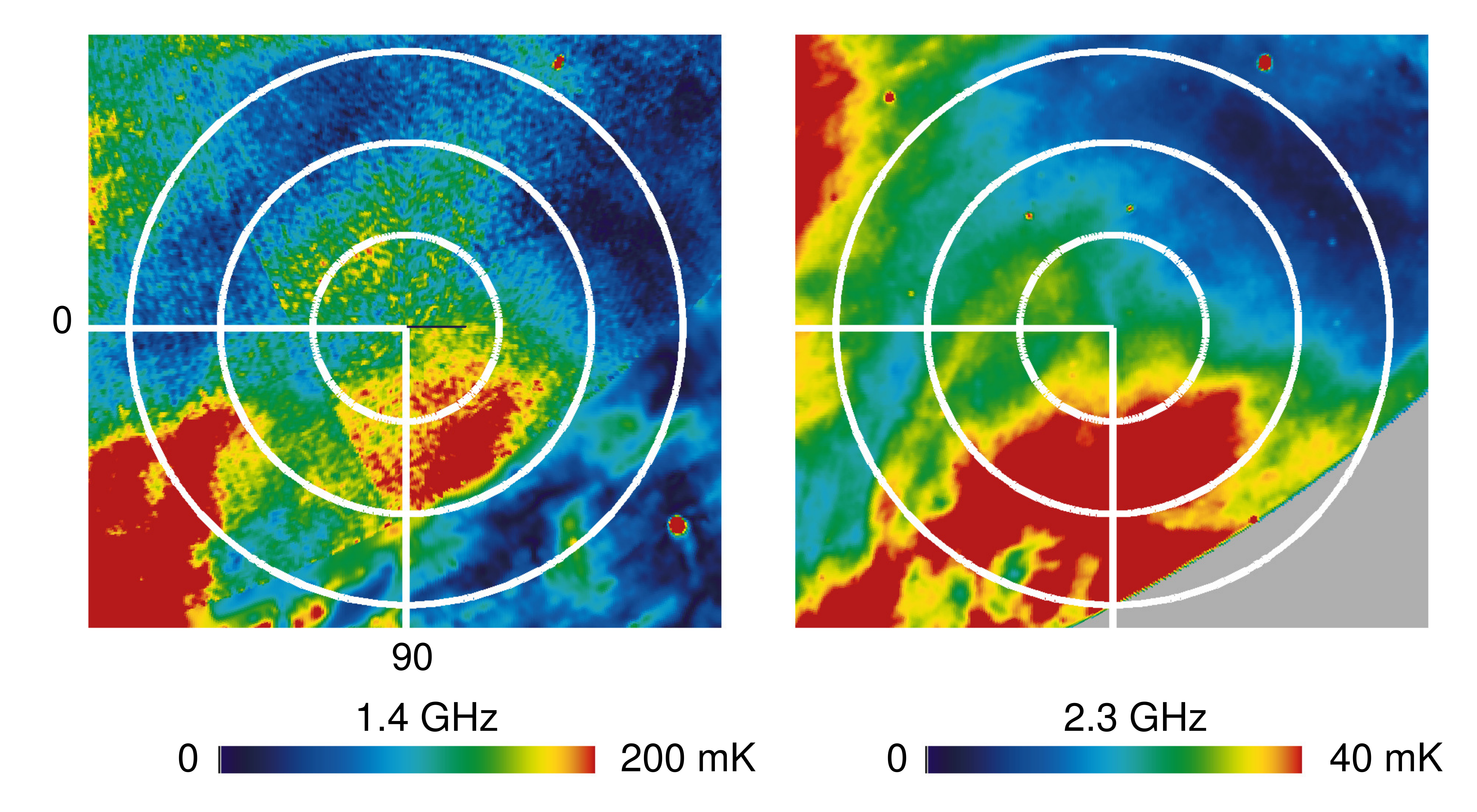}
    \caption{Polarized intensity maps at 1.4 GHz (left, DRAO + Villa Elisa) and 2.3 GHz (right, S-PASS smoothed to the resolution of the 1.4 GHz observations).
    Projections are centered on the south Galactic pole, with white circles depicting latitudes  
    $b = 80^\circ$, $70^\circ$ and $60^\circ$ moving outward from the pole. 
    Galactic longitudes $l=0^\circ$ and $90^\circ$ correspond to the horizontal and vertical white lines respectively.
    Regions with no survey coverage are shown in gray. 
    While Faraday effects are minimized for $b < -70^\circ$, visible scan striping variations and discontinuities
    are present in both surveys contributing to the 1.4~GHz map.
    (There is an unrelated map processing artifact near the pole at longitude $180^\circ$). 
    These effects act to reduce the spatial correlation between the surveys and
    inhibit construction of a spatially detailed spectral index map between the two frequencies.}
    \label{fig:southpole_im}
\end{figure*}

\wmap\ and \planck\ data in the $20 - 70$~GHz range provide high quality all-sky maps of the polarized synchrotron signal (with subdominant additional contributions from CMB and thermal dust components).
Instrument noise and the relatively steep decrease in synchrotron signal with increasing frequency combine
to cause the SNR of these maps at high latitudes to be lower than needed to produce tightly constrained 
maps of $\beta^{pol}$ on few-degree (or smaller) scales \citep{planck/04:2018}.  Numerous investigations have produced \wmap-only,
\planck-only, or combined \wmap\ and \planck\ estimates on larger patches (e.g. \citealt{kogut/etal:2007, dunkley/etal:2009jp, fuskeland/etal:2014, choi/page:2015, planck/11:2018, jew/grumitt:2020, beyondplanck/fg:2020,
martire/etal:2021}).  Methods have included
both pixel-space and harmonic-space (power-spectrum) based techniques.
As an example of the uncertainties achieved using up to 71\% of the sky excluding planar regions, and the full range of \wmap\ and \planck\
frequencies in a power spectrum analysis, \citet{planck/11:2018} computed a weighted mean and standard deviation of
$\beta^{pol} = -3.13 \pm 0.13$.

Supplementing \wmap\ and/or \planck\ observations with data at lower ($< 20$~GHz) frequencies 
is one option for
production of a finer spatial resolution map of $\beta^{pol}$, as lower frequency observations have the advantage of a brighter intrinsic synchrotron signal.  There are however,
challenges: (1) depending on the frequency, Faraday rotation and depolarization along the LOS can be
substantial, (2) for ground-based or balloon experiments, full-sky coverage is the exception, and 
calibration challenges are introduced from the atmospheric and ground environment, and (3) a greater 
frequency ``lever-arm'' from target cosmological frequencies increases the error arising from
potential spectral curvature deviations from a pure power-law dependence.

Another analysis option is to determine $\beta^{pol}$ using only lower frequency data ($\nu < 20$ GHz).
Although the synchrotron signal is considerably higher at these frequencies, the challenges listed above
also apply in this case.  Absolute gain calibration uncertainties of order 5\% can also play a larger 
role in this case than when the frequency lever arm is longer.

At present, the only publicly available lower frequency polarization data with substantial sky coverage
are those at 1.4 and 2.3~GHz. In the following subsections we discuss high-latitude $\beta^{pol}$ results obtained
from three different combinations of frequency bands: (1) 1.4 and 2.3 GHz, (2) 2.3, 23 and 33~GHz, and
(3) 1.4, 23 and 33~GHz.  We analyze frequencies $\lesssim 40$~GHz to avoid the increasing fractional contribution of CMB and dust components at higher frequencies, and omit \planck\ 30 GHz in order to simplify our search for systematic signatures in fit residuals. 

\subsection{1.4 and 2.3 GHz } \label{sec:offplane_radio}

\begin{figure}
    \centering
    \includegraphics[width=3.5in]{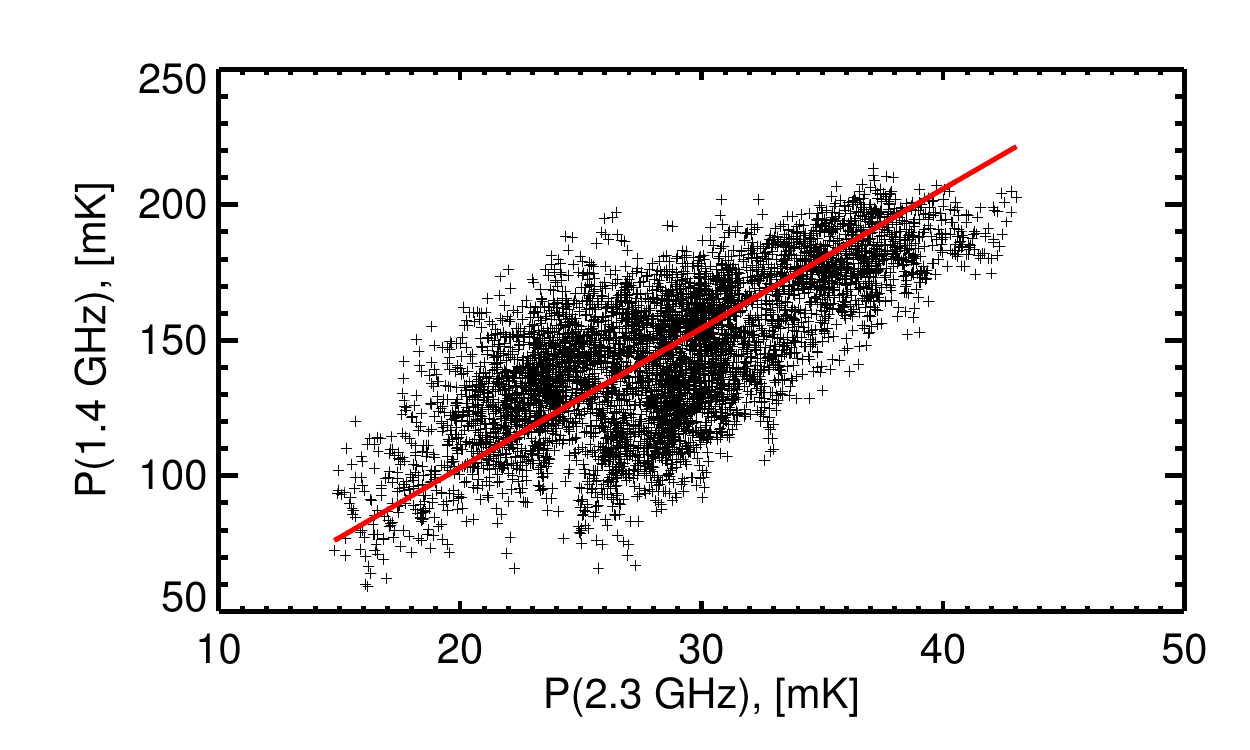}
    \caption{Determination of $\beta^{pol}$ between 1.4 and 2.3 GHz, using the correlation slope method.
    Data (black points) correspond to polarized intensities $P$ within a $10^\circ$ radius of the  Galactic south pole
    at the resolution of the 1.4~GHz map.  The fit to the slope (red line) corresponds to $\beta^{pol} = -3.30 \pm 0.03$.  The scatter in the plot reflects the pattern differences seen in Figure~\ref{fig:southpole_im}.
    }
    \label{fig:southpole_tt}
\end{figure}

The merged full-sky 1.4~GHz DRAO and Villa Elisa surveys overlap the footprint
of the southern S-PASS 2.3~GHz survey.  However, Faraday depolarization effects are
more significant at 1.4~GHz than at 2.3~GHz, limiting the useful sky overlap between the two surveys
to those portions for which $|b| \gtrsim 45^\circ$ \citep{carretti/etal:2010}. 
A further complexity is introduced by the few visually apparent artifacts in the DRAO northern survey
\citep{wolleben/etal:2006} and scan striping in the southern Villa Elisa 1.4~GHz data \citep{testori/etal:2008}.
Figure~\ref{fig:southpole_im} compares the polarized intensity $P$ ($=\sqrt{Q^2 + U^2}$) at the
two frequencies for a cap about the southern Galactic pole.   The 2.3 GHz survey data on the right
has been smoothed to match the $\sim 36'$ FWHM of the 1.4 GHz map on the left. 
Since instrumental noise is low compared
to the signal, noise bias in $P$ is not prominent at either frequency, and the ``noise'' seen in the
1.4~GHz map is calibration-related.  \citet{testori/etal:2008} note that the scan striping smooths
out over scales of several degrees.  Visual examination indicates the region shown presents
the 1.4~GHz surveys at their worst in terms of these calibration-related features.  
We will discuss further consequences of these features in Section~\ref{sec:offplane_3freq}.

Figure~\ref{fig:southpole_tt} shows a correlation plot
between 1.4 and 2.3~GHz polarized intensity
data for $b \leq -80^\circ$.  
We have used P rather than Q and U in this specific case because 
the high SNR mitigates against noise bias concerns normally present when working with P
and because the data are not sufficiently constraining to independently solve for $RM$.
The correlation slope obtained using statistical uncertainties corresponds to $\beta^{pol} = -3.30 \pm 0.03$. This is a value consistent with other analyses, but the scan striping noise requires use of larger sky areas
than the native survey resolution.  Furthermore, the presence of artifacts and gain 
discontinuities can skew
slope determinations.  The impact is higher for determinations using shorter frequency lever arms.
For example, a 5\% gain variation at 1.4~GHz translates to an error in $\beta^{pol}$ of
$\sim 0.1$ if the second frequency is 2.3~GHz, but only $\sim 0.02$ if the second frequency is 23~GHz.
It is more advantageous to proceed with
computing separate fits of [1.4, 23, 33] and [2.3, 23, 33] at a lower spatial resolution and to compare  results for those pixels in common.

\subsection{[1.4, 23, 33] and [2.3, 23, 33] Fits} \label{sec:offplane_3freq}

There are results in the literature that use combinations of 1.4~GHz and \wmap\
frequencies, and separately 2.3~GHz combined with \wmap\ and/or \planck\ LFI data.
\citet{carretti/etal:2010}  determined a mean full-sky value of $\beta^{pol} = -3.2$ using
the combined DRAO and Villa Elisa 1.4~GHz data and an older \wmap\ 5th-year release 23~GHz map, but provided no commentary on uncertainties.
\citet{krachmalnicoff/etal:2018} analyzed 2.3, 23, and 30~GHz 
polarized intensity maps with a Galactic latitude cut $|b|< 20^\circ$ to avoid regions affected by Faraday depolarization.
On $2^\circ$ scales, they determined $-4.4 \le \beta^{pol} \le -2.5$, but with
only 12\% of the sample found to differ at high significance from the mean of $-3.2$.
\cite{fuskeland/etal:2019} used a correlative analysis between \wmap\ 23~GHz and S-PASS to determine $\beta^{pol}$ within $15^\circ$ patches in the southern sky.  They applied Faraday rotation corrections to the S-PASS Q and U maps, choosing the 
S-PASS, \wmap\ and \planck-based RM map of \cite{carretti/etal:2019} over an older extragalactic source-based map \cite{hutschenreuter/ensslin:2020}.

\begin{figure*}[ht]
    \centering
    \includegraphics[width=6.5in]{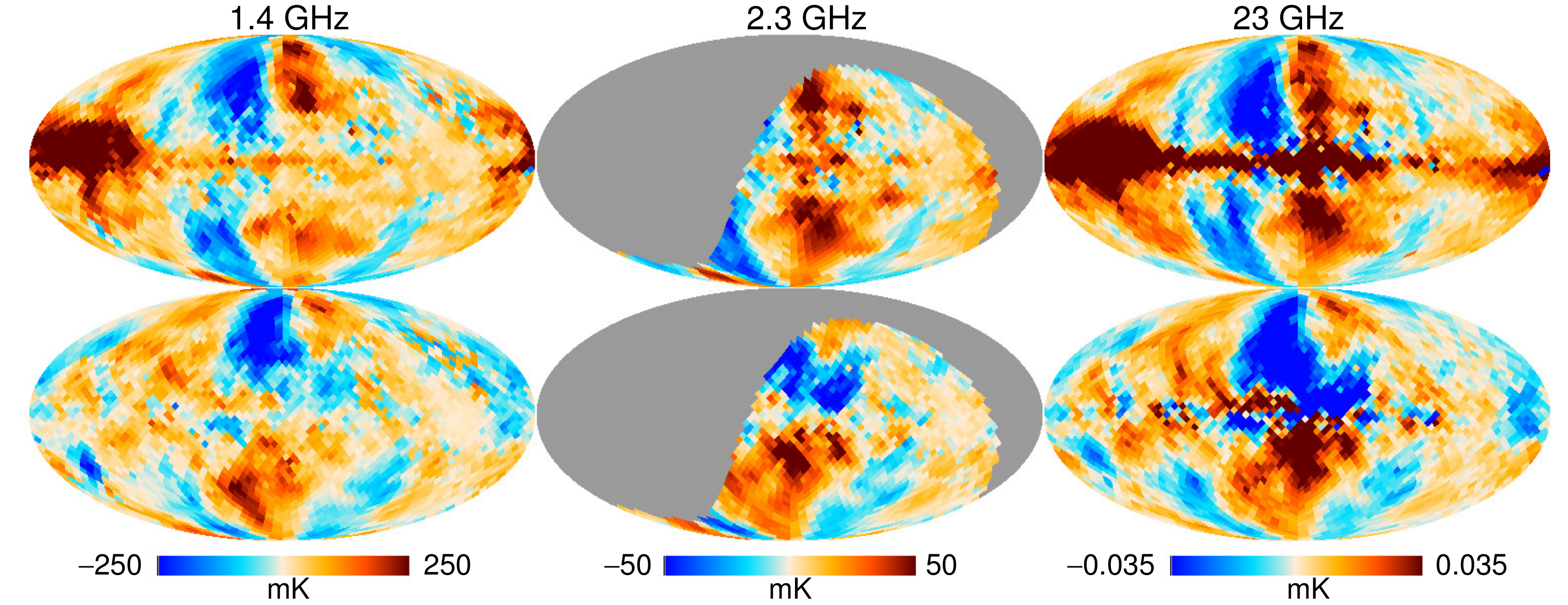}
    \caption{ {\it{Left:}} The 1.4~GHz Stokes $Q_0$ (top) and $U_0$ (bottom) parameter maps resulting from the [1.4, 23, 33] fit.  The $Q_0$ and $U_0$ maps are
    corrected for Faraday rotation as determined from the $RM$ fit parameter (equation 2).  {\it{Middle: }} The 2.3~GHz $Q_0$ and $U_0$ maps from the [2.3, 23, 33] fit.
    The portion of the sky unobserved by the 2.3~GHz S-PASS survey is shown in gray.  {\it{Right:}} 
    The observed \wmap\ 23~GHz (K-band) Q and U maps  
    are shown for comparison.  Scale stretches are chosen to match a 
    $\nu ^{-3.2}$ power law.  Close to the Galactic plane, the synchrotron polarization direction is nearly perpendicular to the plane, resulting in stronger Stokes Q 23~GHz emission there compared to Stokes U. (see e.g., Fig 6 of \citealt{bennett/etal:2013}).
    Regions of depolarization in the vicinity of the Galactic plane are indicated in the 1.4 and 2.3 GHz maps by the signal deficit compared to 23~GHz. 
    }
    \label{fig:q0u0}
\end{figure*}

\vspace{3mm}
\subsubsection{Fitting Method}

Our fits to [1.4, 23, 33] and [2.3, 23, 33] frequency combinations have similar analysis elements
to those described above.
However, there are differences in details, including chosen spatial resolution,
sky fraction, selected frequencies, and treatment of Faraday rotation.
We work at HEALPix $N_{side}=16$, or roughly a pixel resolution of $3.7^\circ$.  Working at this 
spatial scale increases the SNR of the polarization data over native resolution and allows us to use other products that are available at the same resolution, in particular covariance matrices and the \wmap\ loss imbalance templates. 

The following model is fit to the Q and U data in pixel-space for the three frequencies:
\begin{align}
      [Q, U]^{obs}_{\nu_0} &  =  {{f}} (RM, Q_0, U_0) \\
      [Q, U]^{obs}_{23}     &  =  1.014~[Q_0, U_0]~(22.5/\nu_0)^{\beta^{pol}} \\
      [Q, U]^{obs}_{33}    &  =  1.026~[Q_0, U_0]~(32.65/\nu_0)^{\beta^{pol}} 
\end{align}
where $\nu_0$ is either 1.41 or 2.303~GHz, and the
free parameters $RM$, $Q_0$, $U_0$, $\beta^{pol}$ are evaluated for each $N_{side}=16$ pixel.
$Q_0$ and $U_0$ are the intrinsic (unrotated) Stokes parameter maps at frequency $\nu_0$.
The function ${f}$ transforms the rotation measure parameter to a
rotation angle $\theta$ assuming a $\lambda^{-2}$ dependence and
applies the corresponding rotation matrix (see Section~\ref{sec:rm_data}) to the intrinsic synchrotron amplitudes
$Q_0$ and ${U_0}$ to model the observed 1.4 or 2.3~GHz Q and U maps.  Faraday rotation is ignored for 23 and 33 GHz bands. 
However, the 23 and 33 GHz maps are in thermodynamic temperature units, so the factors 1.014 and 1.026
provide the conversion from model RJ units to thermodynamic \citep{bennett/etal:2013}.
The model assumes
the spectral index has no frequency dependence.

We use the scipy non-linear fitting routine {{\textsl{curvefit}}} for parameter estimation with weighting specified 
by QQ, QU and UU variances per pixel for each frequency (there are no QU variances available at 1.4 or 2.3 GHz).  The Faraday depth map of \citet{hutschenreuter/etal:2021} is used for the
initial $RM$ guess to the iterative fit.

Simulations are used to help understand errors and potential systematic effects in the real data.
We generate simulated $N_{side}=16$ sky maps containing CMB, synchrotron and instrument noise
at the same frequencies.  We use the same sky signal model as that described in Section~\ref{sec:hisig}. 
The simulated 2.3~GHz noise is based on the delivered Q and U noise maps, and the instrument noise for 1.4~GHz is generated based on the single value provided for each survey.
Pixel-pixel covariance matrices for \wmap\ 23  and 33~GHz bands are used to generate noise 
realizations for these frequencies.  Use of the full pixel-pixel covariance incorporates the
off-diagonal terms and large-scale modes of enhanced uncertainty discussed in Section~\ref{sec:data}.
Faraday rotation under the assumption of thin-screen conditions is included in
the simulation of the 1.4 and 2.3~GHz synchrotron emission, using the Faraday depth map of \citet{hutschenreuter/etal:2021} (see Section~\ref{sec:rm_data}). 
As described in the next section, we do not analyze regions expected to strongly violate the thin-screen approximation.

\begin{figure*}    
    \centering
    \includegraphics[width=6.5in]{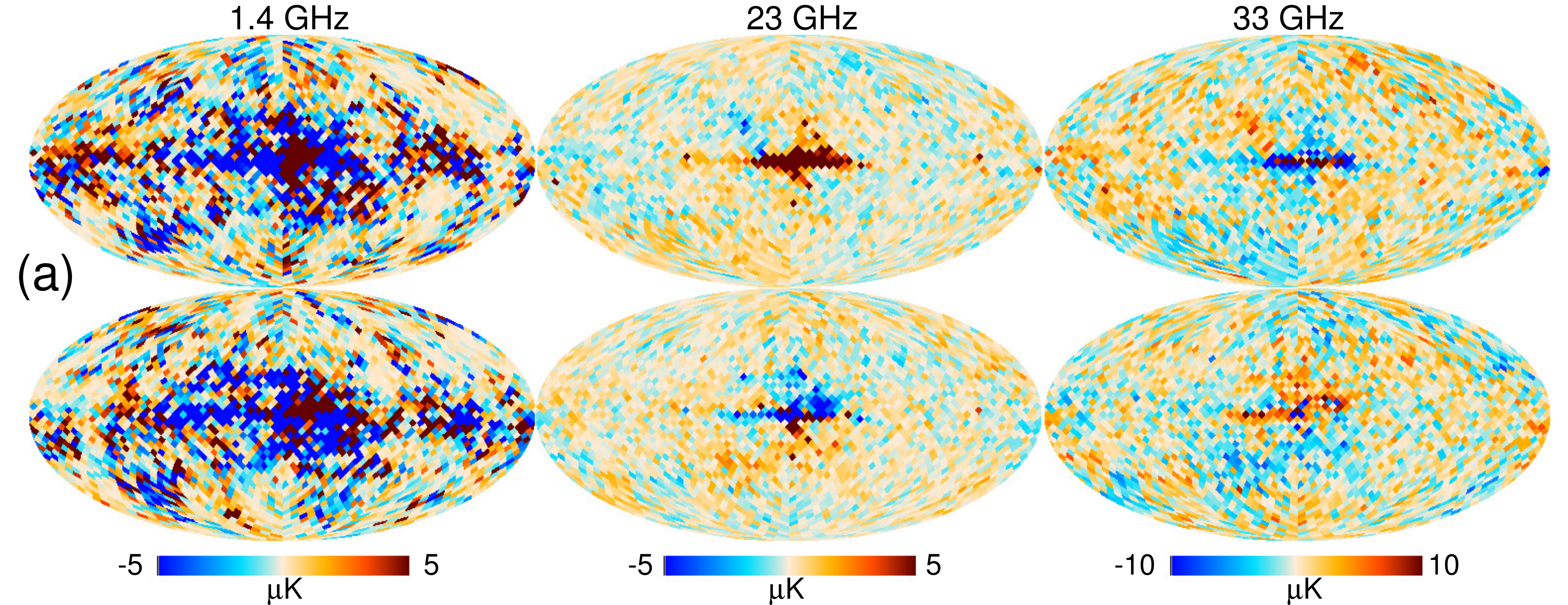}
    \par
    \vspace{3mm}
    \includegraphics[width=6.5in]{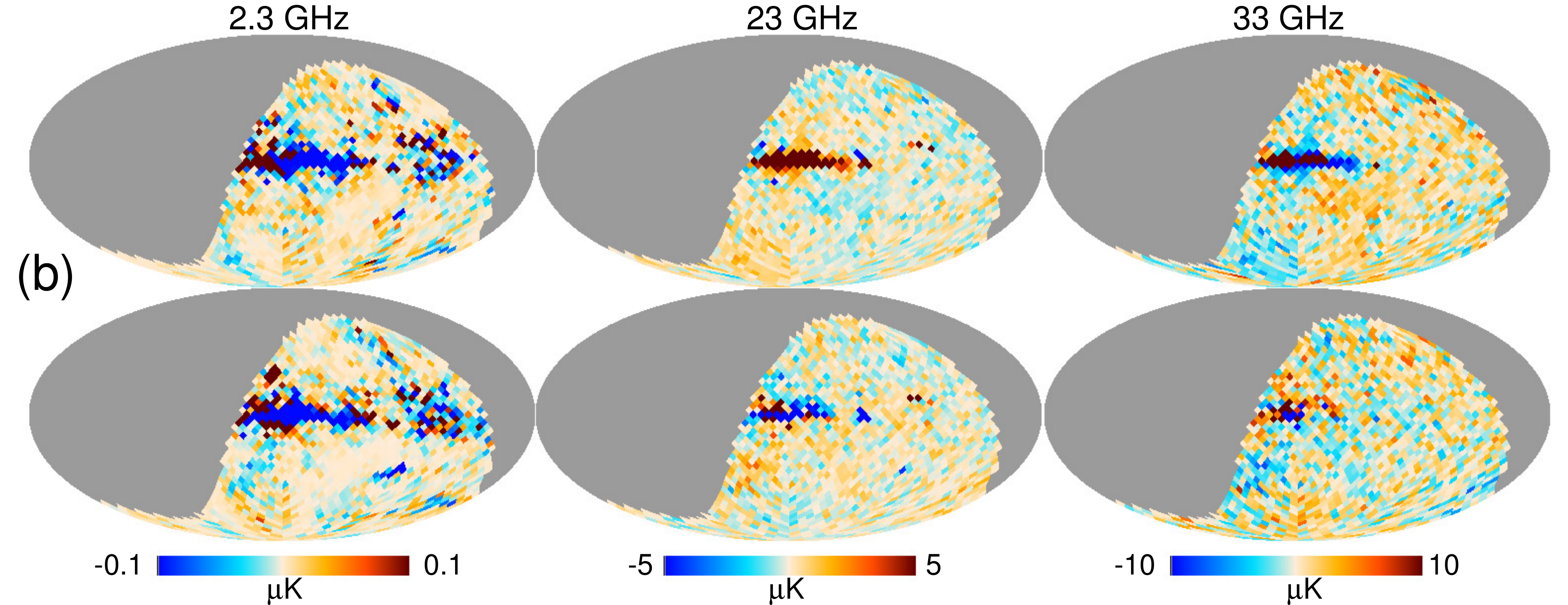}
    \caption{ {\it{(a)}:} Residual (data minus model) Stokes Q (top) and U (bottom) maps for the [1.4, 23, 33] frequency combination.
    {\it{(b):}} As with {\it{(a)}}, but residual maps are for the [2.3, 23, 33] frequency combination.
    Residuals at 23 and 33~GHz are visually dominated by instrument noise in most locations.  However, poor fits
    resulting from strong depolarization in the low-frequency channel
    produce noticeable systematics in the residuals for all frequencies.
    Systematics from depolarization effects extend up to $|b| \sim 45^\circ$ in the [1.4, 23, 33] residuals and are particularly clear as
    dark red and blue shadings in the 1.4~GHz residuals.
    Pixel masking criteria include rejection of outliers in the low frequency residual maps.
    }
    \label{fig:fit_resids}
\end{figure*}

\subsubsection{Pixel Selection}

We fit the parameterized model described in the previous section to all available pixels: full sky for the [1.4, 23, 33] frequency set
and the S-PASS southern footprint for the [2.3, 23, 33] fit.  We then apply a set of pixel quality
masks based on criteria that we derive from fit residuals, parameter errors, and estimates of
potential systematic effects.   

Figure~\ref{fig:q0u0} shows the Faraday rotation corrected amplitudes $Q_0$ and $U_0$ fit for 1.4~GHz (left column) and
2.3~GHz (middle), in comparison with the observed \wmap\ 23~GHz maps (right).  Display scales are chosen
such that if the sky synchrotron signal behaved as $\nu^{-3.2}$ over the entire sky, the Q and U maps
at all three frequencies would look the same, which they do not.  The clearest visual difference
takes the form of weaker emission near the Galactic plane for the 1.4 and 2.3~GHz maps compared to
23~GHz.  This is the result of Faraday depolarization in these regions, which causes a shallower
spectral index.  This is an example of pixels that should be masked from the fits because of a
systematic effect, but there are additional reasons for masking.

The following criteria are applied to mask pixels.  The masks are not necessarily exclusive of
each other, in that some effects show up in more that one set of criteria:

(1) Pixels containing emission from strong polarized sources are masked.  This includes extragalactic sources
such as Fornax A, Cen A and 3C279.  In many cases, the source masking was not necessary because
those pixels were masked by other criteria as well.

(2) Outliers in the low frequency fit residual maps are masked.  Figure~\ref{fig:fit_resids} shows
the residuals (data minus model) for both the (a) [1.4, 23, 33] and (b) [2.3, 23, 33] fits.  
Compared to the 23 and 33~GHz data, the relative SNR for 1.4 and 2.3~GHz is very high and the fit
essentially treats these frequencies as noiseless.  
This causes the $Q_0$ and $U_0$ parameters to be highly dominated by the Q and U maps
of the lower frequencies, modulo the rotation measure correction.  The residuals at these frequencies
are below noise level, but show some structure. In particular, outliers in the 1.4~GHz and 2.3~GHz
residual maps
coincide with areas of strong disgreement between e.g. [1.4, 23] and [23, 33].  This includes regions
of strong depolarization and potential map artifacts.

\begin{figure}[tb]
    \centering
    \includegraphics[width=2.5in]{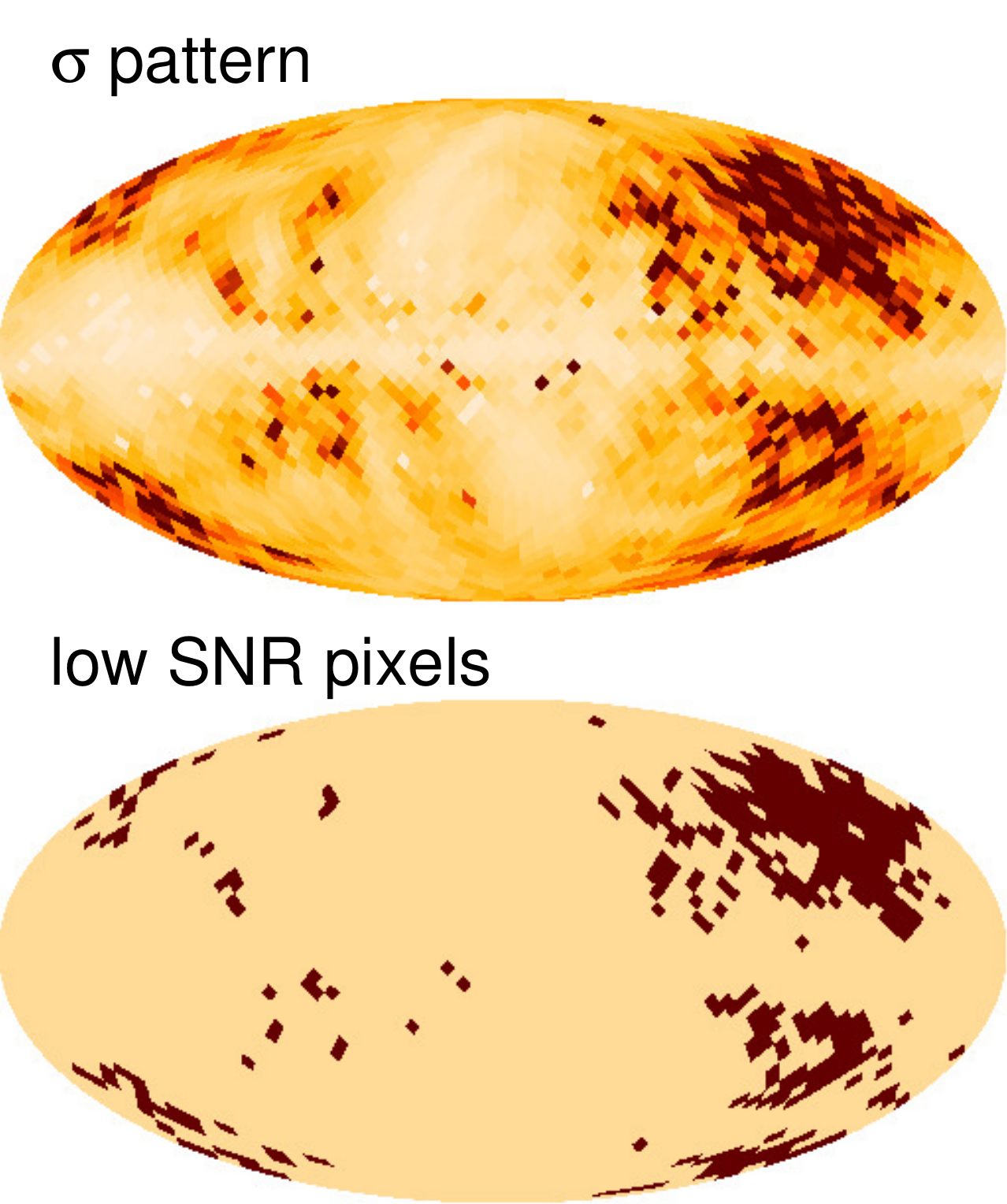}
    \caption{Given the comparatively higher SNR at 1.4 and 2.3 GHz, the statistical uncertainties in the [1.4, 23, 33] and [2.3, 23, 33] parameter fits are primarily governed by the 23~GHz SNR.  {\it{Top: }}The statistical uncertainty $\sigma (\beta^{pol})$ from the [1.4, 23, 33] fit, scaled so that $\sigma (\beta^{pol}) \geq 0.2$ saturates at a dark brown color.  
    {\it{Bottom: }}Mask showing in brown those $N_{side}=16$ pixels for which the 23~GHz polarized intensity is at the level of instrument noise (all others orange).
      Since parameters derived for these pixels are not well constrained, we exclude these
    locations from the parameter maps.
    }
    \label{fig:stat_unc}
\end{figure}

(3) Pixels with high statistical uncertainties in $\beta^{pol}$ are masked.  The statistical
uncertainties in fit parameters $\beta^{pol}$ and $RM$ follow the same pattern, illustrated in
the top panel of Figure~\ref{fig:stat_unc}.  This pattern strongly correlates with the SNR of the 
23~GHz
polarized intensity.  This is expected given the relative SNR of the low frequency in each 3-frequency
set, and because the 33~GHz band has a lower SNR than that of 23~GHz.  
Simulations show the same effect, but
because the simulated sky model is not identical to the data, the precise pattern is not replicated.
Instead, the bottom panel of Figure~\ref{fig:stat_unc} shows our estimate of the lowest SNR pixels
in $P$ at 23~GHz.  The estimate is based on evaluating $P^2$ from data splits and noting those pixels that
can achieve negative values.  We mask pixels for which $\sigma (\beta^{pol}) > 0.2$.

(4) Pixels likely affected by Faraday depolarization are masked.  We employ both a Galactic latitude
cut and a cut on Faraday depth.  The latitude cuts are $|b| < 40^\circ$ for the [1.4,~23,~33] fits,
and $|b| < 15^\circ$ for the [2.3,~23,~33] combination.  For both frequency combinations, pixels with
Faraday depth $\phi > 40$ rad m$^{-2}$ are excluded.  
This value was chosen as an approximate threshold where the values of $\phi$ and $RM$ shown in
Figure~\ref{fig:rm_maps} are in agreement within uncertainties, thus selecting lines of 
sight where the thin screen assumption is most likely to be valid, and depolarization of least concern.

\subsubsection{Comparison of [1.4,23,33] and [2.3,23,33] results}

In this section, we discuss systematic spatial differences between the $\beta^{pol}$ and $RM$ parameter maps obtained 
from separate fits to [1.4, 23, 33] and [2.3, 23, 33] GHz. Since both of
the three frequency fits each share the same 23 and 33 GHz data, differences in recovered
parameters must result from either the model fit assumptions as a function of frequency (e.g. no curvature in the spectral index) 
or the low frequency data themselves (e.g., calibration systematics).
For example, the three-frequency model described by equations 2, 3, and 4 of section 4.2.1 computes a Faraday rotation term
only for the lowest frequency. The model fit adjusts the $RM$ value in each pixel to enforce agreement with 
the 23 GHz polarization angle (with 33 GHz subdominant because of its lower SNR).  Artifacts in the 
lowest frequency Q and U maps affecting polarization angle will result in an error in the recovered $RM$ parameter, 
and thus affect the recovered $Q_0$ and $U_0$ values as well.  Artifacts affecting the polarized intensity will bias
the recovered value of  $\beta^{pol}$.  The two effects
are not necessarily mutually exclusive.

\begin{figure}[t]
    \centering
    \includegraphics[width=3.25in]{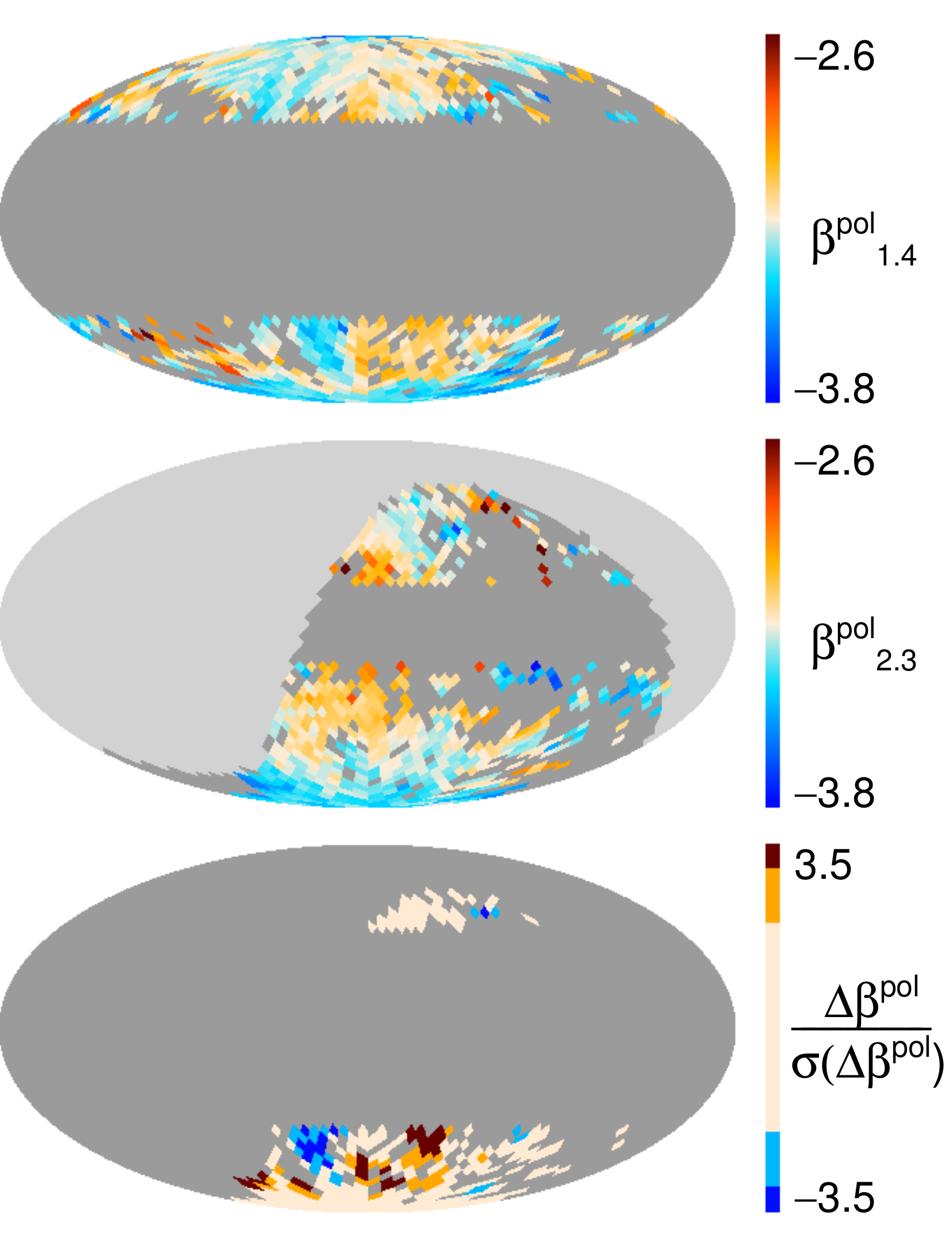}
    \caption{Polarized spectral index results for individual fits after pixel selection.
    {\it{Top:}} $\beta^{pol}$ for [1.4, 23, 33] GHz. {\it{Middle:}} $\beta^{pol}$ for [2.3, 23, 33] GHz.
    {\it{Bottom:}} The difference between the top and middle maps, divided by its 1-$\sigma$ uncertainty.
    This is shown only for pixels in common between top and middle maps.
    There are clusters of pixels in the south for which 
    {${\Delta} {\beta}^{pol} /{\sigma}({\Delta} {\beta}^{pol})$} 
    indicates significant disagreement. The blue region in the south near $l \sim 30^\circ$
    is associated with a particular 1.4~GHz feature.
    }
    \label{fig:beta_results}
\end{figure}

\begin{figure}[t]
    \centering
    \includegraphics[width=3.25in]{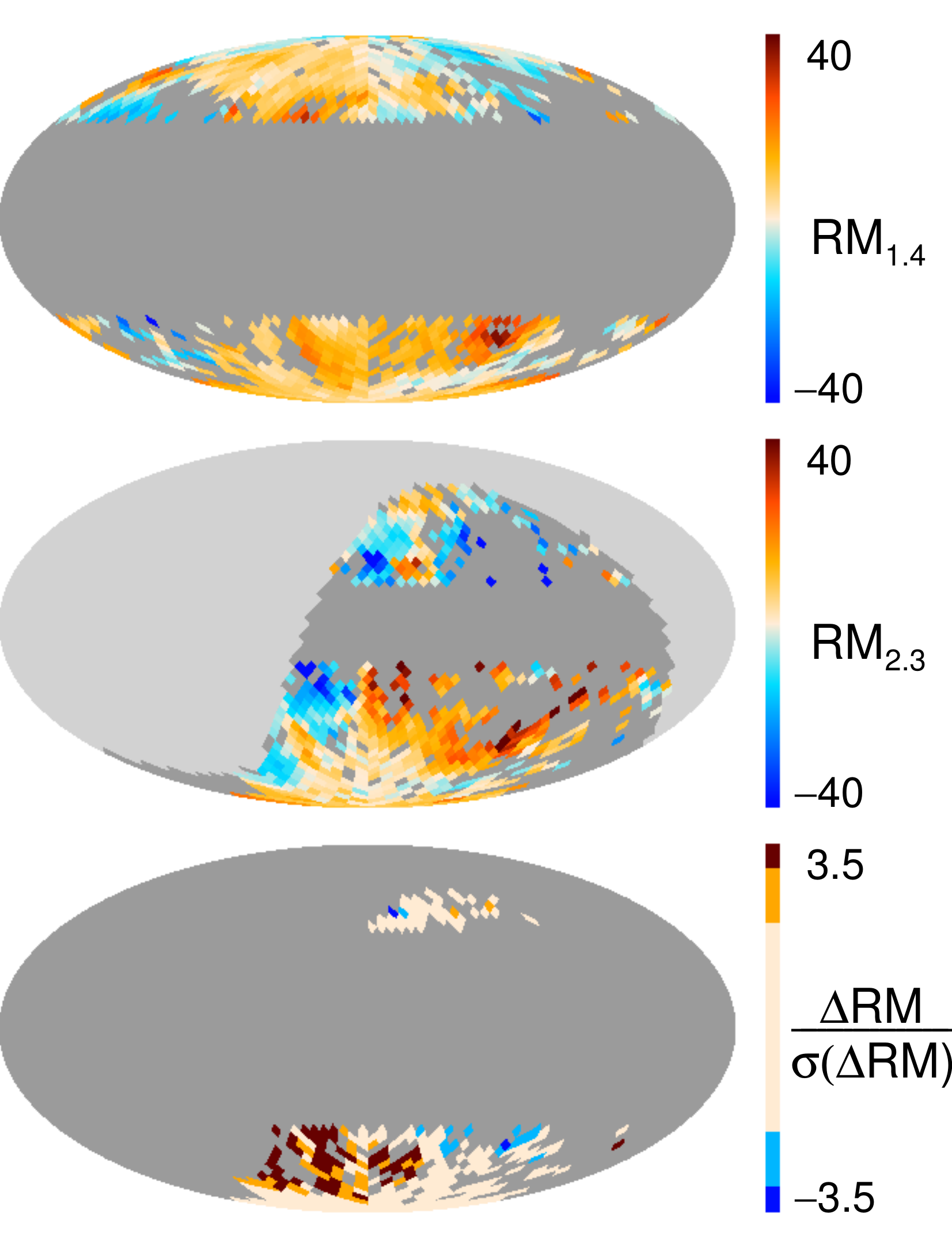}
    \caption{Rotation measure results for individual fits after pixel selection.
    {\it{Top:}} RM from fit to [1.4, 23, 33] GHz, in units rad m$^{-2}$. {\it{Middle:}} RM from fit to [2.3, 23, 33] GHz, same units.
    {\it{Bottom:}} The difference between the top and middle maps, divided by its 1$\sigma$ uncertainty.
    This is shown only for pixels in common between top and middle maps.
    There are pixel clusters in the south for which 
    {${|\Delta} RM| /{\sigma(\Delta} RM)$} indicates significant disagreemnt. 
    These clusters occur in the same general location as those pixels for which 
    {${|\Delta \beta^{pol}| / \sigma(\Delta \beta^{pol})}$} is high (see 
    Figure~\ref{fig:beta_results}).
    }
    \label{fig:rm_results}
\end{figure}

In Figure~\ref{fig:beta_results}, we show the $\beta^{pol}$ parameter maps from each of the 3-frequency fits, and in 
Figure~\ref{fig:rm_results} we show results for the $RM$ parameter.  
In both figures, the top plot corresponds to results from
the [1.4, 2 3, 33] fit, and the middle plot shows those for [2.3, 23, 33].  
The bottom plot in each panel of Figures~\ref{fig:beta_results} and \ref{fig:rm_results} shows 
the per-pixel difference between the [1.4, 22, 33] and [2.3, 22, 33] parameter maps, and expresses the differences $\Delta$ as a 
fraction of an individual parameter's uncertainty $\sigma(\Delta)$. The use of $\sigma(\Delta)$ here is not precise, 
as we use the root sum square of the uncertainties from the two fits even thought we are not dealing with completely independent datasets.
However, it is a useful representation in terms of magnitude of deviation.
These ${\Delta/\sigma(\Delta)}$ maps have been binned with color demarcations at [$-3.5, -3.0, -2.5, -2.0, 2.0, 2.5, 3.0, 3.5$]. 
Pixels with $\Delta$/$\sigma(\Delta)$
that exceed $-3$ $(3)$ show as dark blue (dark brown).
In the case of the $\beta^{pol}$ uncertainty, the statistical uncertainty derived from the fit
has been summed in quadrature with an additional 0.02 uncertainty that assumes an absolute gain uncertainty of 5\%
\citep{carretti/etal:2019}.  

In these figures, pixels with the largest 
disagreement in the {${\Delta/\sigma(\Delta)}$} maps
are spatially clustered, rather than randomly scattered over the sky.  The cluster of blue
pixels in the {${\Delta\beta^{pol}/\sigma(\Delta\beta^{pol})}$} map near $l \sim 30^\circ$, $45^\circ < b < 70^\circ$ is associated with a region of high 
polarized intensity emission in the 1.4 GHz map which is not well correlated with emission in
the 2.3~GHz map. The feature can been seen as the region of red in the lower left corner of the left panel of Figure~\ref{fig:southpole_im} and also most easily seen in the $U_0$ map of Figure~\ref{fig:q0u0}.
This roughly wedge-shaped region lies in the declination overlap strip 
($-29^\circ < \delta < -10^\circ)$ between the DRAO and Villa Elisa 1.4~GHz surveys.  We have been 
showing results from the merged 1.4~GHz survey map, but also performed the same parameter model fit for the individual surveys
to determine if there were significant differences in the parameter results in this region.
Although there are clear differences in smaller-scale morphology and intensity, both surveys see the same general
feature, which produces a steeper $\beta^{pol}$ and higher $RM$ in this region than seen at 2.3 GHz.
A bright feature is clearly not the result of depolarization, nor does it fit with the physical
picture of synchrotron spectral index flattening with decreasing frequency.  
However, \citet{wolleben/etal:2006} note the possibility of features from uncorrected beam sidelobes 
in the DRAO survey.  The $RM$ computed
for this region echoes the wedge-shape, and is not as morphologically consistent with the
Faraday depth map as the $RM$ derived from the 2.3~GHz data in the same location.  For this region,
evidence points to the 1.4 GHz map as the less accurate observation.  Given the
known presence of artifacts in the 1.4 GHz maps, it is likely that the 1.4 GHz data are the origin of
the remaining discrepant pixels as well.

The above discussion raises the question of the accuracy of the northern portion of the [1.4, 23, 33] parameter maps,
where there is as yet no complementary survey at a similar frequency for comparison.  In this case, we must rely
on secondary indicators of potentially compromised regions: general (but not exact) morphological agreement with the Faraday
depth map, and visual detection of features in the $Q$ and $U$ maps which appear unusual compared to those of 23 and
33 GHz.  Guided by the types of effects seen in the 1.4 and 2.3 GHz comparison, we enlarged the pixel exclusion radius around one
particular feature in the [1.4, 23, 33] residual map centered around $l \sim 100^\circ, b \sim -50^\circ$ (see Figure~\ref{fig:fit_resids}).
We did not find grounds to suspect additional regions of the 1.4 GHz survey 
for which $\delta \geq 0$, and thus proceeded with our analysis.

Most of this section's discussion has centered on a comparison of the [1.4, 23, 33] and [2.3, 23, 33] results, but
we also compare the [2.3, 23, 33] results with those previously published.
For those pixels meeting the selection criteria described in Section 4.2.2, the  [2.3, 23, 33] $\beta^{pol}$ map presented here
is consistent with those of \citet{krachmalnicoff/etal:2018} and \citet{fuskeland/etal:2019}.

\begin{figure}[ht]
    \centering
    \includegraphics[width=3.5in]{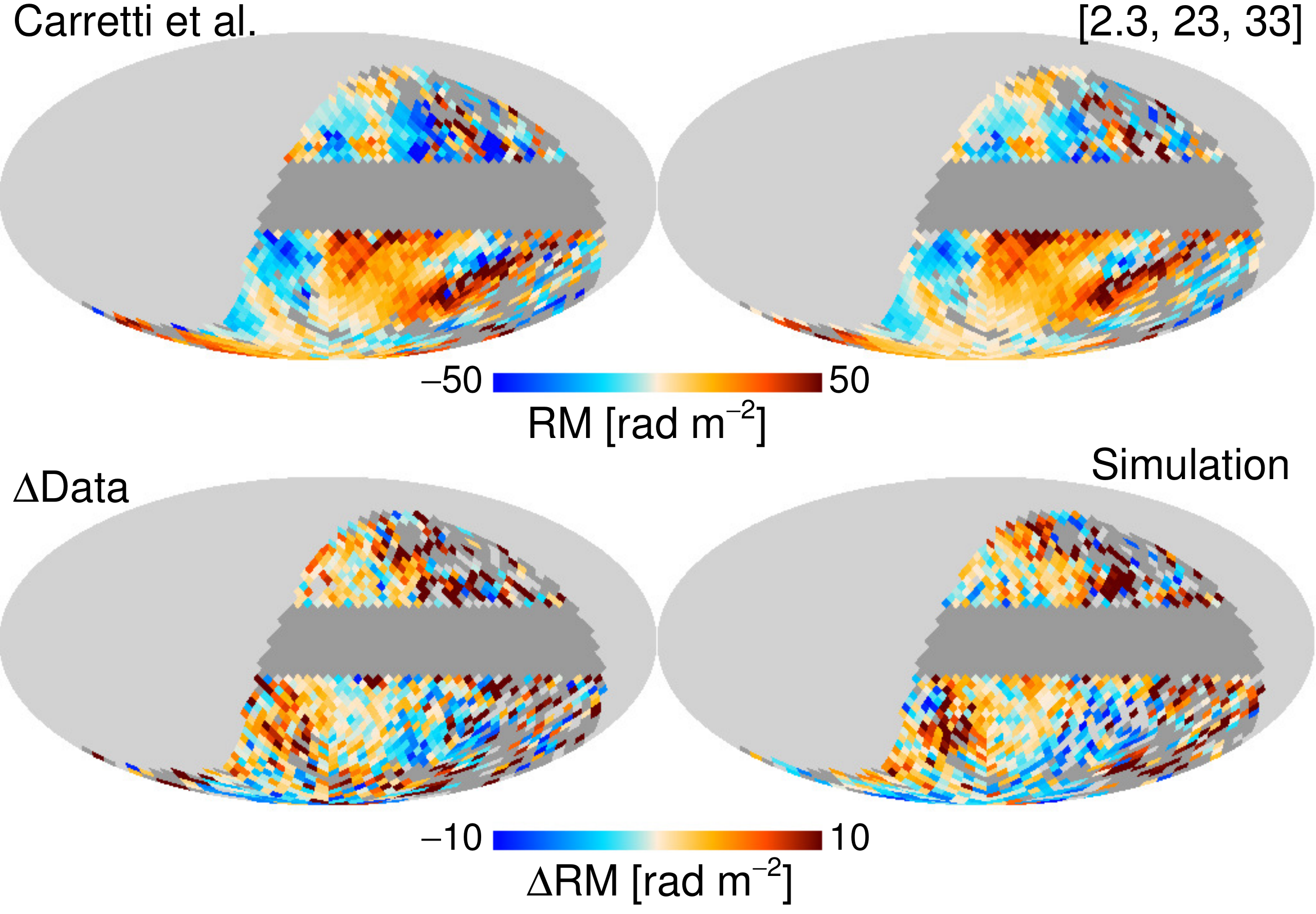}
    \caption{A comparison of $RM$ values derived from S-PASS 2.3 GHz data in combination
    with either a mix of \planck\ and \wmap\ data, or \wmap\ data alone.{\it{Top left}:} $RM$ map derived by
    \citet{carretti/etal:2019} using 2.3, \wmap\ 23 and LFI 30 GHz data, degraded to $N_{side}=16$.
    {\it{Top right}:} $RM$ map derived in this work from 2.3, \wmap\ 23 and \wmap\ 33 GHz data at the same
    resolution.  Masked pixels
    are shown in dark gray, but note we show a larger set of pixels than used in our main analysis,
    including some higher noise regions. (see text).  {\it{Bottom left}:} The top right
    [2.3, 23, 33] values minus the top left map of \citet{carretti/etal:2019}.
    There are systematic large-scale differences at a low level. 
    {\it{Bottom right}:} The same fitting technique used for the data is applied to simulated 
    frequency combinations of  [2.3, 23, 33] and [2.3, 23, 30, 33] GHz.  A large-scale
    systematic based on 30 GHz - 0.47$\times$23 GHz Q ad U differences has been added to the simulated 30 GHz
    map.  This panel shows the difference in the two recovered $RM$ parameter maps, simulating
    the equivalent data map at the bottom left.  The simulation shows that the difference between
    the $RM$ maps derived from the [2.3, 23, 33] fit and that of \citet{carretti/etal:2019} can
    plausibly be attributed to the large scale differences between \wmap\ 23 GHz and LFI 30 GHz.
    }
    \label{fig:rm_vs_carretti}
\end{figure}

We find small differences between the 
$RM$ values we obtained for these selected pixels with those of \citet{carretti/etal:2019} (shown at full resolution in Figure~\ref{fig:rm_maps}).   
At the top left of Figure~\ref{fig:rm_vs_carretti} we show the $RM$ map of \citet{carretti/etal:2019}, degraded from
the original $N_{side}=32$ resolution to $N_{side}=16$; the top right is the $RM$ map we compute
from [2.3, 23, 33].  For both maps, we use a less stringent pixel selection criterion that allows a greater percentage
of sky to be visible. As mentioned in Section~2.2,
\citet{carretti/etal:2019} computed the $RM$ from S-PASS, 23 and 30~GHz data, but noted systematic
differences between 23 and 30~GHz polarization angles that resulted in their choice to exclude pixels for which the 23 and 30 GHz
polarization angles differed by more than $15^\circ$.  In the Figure, we have excluded these same pixels, as well
as pixels within the Galactic Plane subject to depolarization ($|b| < 15^\circ$).  
The two maps in the top row of the Figure look very similar, but 
the difference between them (bottom left) shows indications of low-level large-scale differences.
Since one analysis method includes 30 GHz data whereas the other does not, it is reasonable to posit that
the systematic differences between the LFI and \wmap\ polarization data 
(see Section 3)
are the source of this difference.

We demonstrate that this explanation is likely by creating a synchrotron sky model based on S-PASS polarized intensity,
23 GHz  polarization angles, an assumed constant $\beta^{pol}= -3.2$, nominal instrument noise, and a CMB realization.
Faraday rotation is applied to the simulated 2.3 GHz maps, assuming $RM$ values from the Faraday depth map.
The same model fitting routines used for the data are run on this simulated dataset for three separate combinations: [2.3, 23, 33], 
[2.3, 23, 30, 33] and [2.3, 23, 30$^s$, 33] where the superscript $s$ in the last combination indicates a systematic signature has been added to the 30 GHz Q and U maps.
The systematic signature is computed from the data difference 30 GHz - 0.47$\times$23~GHz (see \citet{weiland/etal:2018}
for further discussion).  

The difference between $RM$ parameters for the [2.3, 23, 33] $-$ [2.3, 23, 30, 33] simulation is
consistent with a null map, whereas the [2.3, 23, 33] $-$ [2.3, 23, 30$^s$, 33] simulated $RM$ map difference,
shown at bottom right of Figure~\ref{fig:rm_vs_carretti}, is very similar to that of the data difference on the
bottom left. 

Since the modeled systematic signature is generated from a difference between LFI and \wmap\ maps, this
by itself does not isolate its origin.
However, 
as discussed in Section 3,
the \planck\ Collaboration associated the dominant contribution to the systematic modes with the PR3 30~GHz maps.
In the next section, we discuss simulations designed to evaluate additional sources of uncertainty.

\subsection{Simulations of potential additional uncertainties} \label{sec:sim_imbal}

Although the large-scale differences between \wmap\ and \planck\ PR3 LFI polarization maps 
are dominated by the LFI gain uncertainty modes, 
how much \WMAP\ contributes to this difference is still a subject of study.

A potential contributing signature in \wmap\ data 
arises from 
uncertainty in the applied correction for imbalance in the transmission efficiencies between the two sides of the differential \wmap\ instrument.  Within the mapmaking framework, an error in the loss imbalance correction 
results in the presence of large-scale 
modes in the \wmap\ maps \citep{jarosik/etal:2003, jarosik/etal:2007, jarosik/etal:2011, bennett/etal:2013}. 
While the spatial modes are well defined by scan geometry, their amplitude is determined by the difference between the
assumed and actual correction values applied to each radiometer.
With two radiometers in each of the 23 and 33~GHz bands, the amplitude of the resultant large-scale 
morphology is dependent on the signed uncertainty for each radiometer.
Loss-imbalance related
modes may be suppressed through the use of the full covariance matrix available for this purpose 
or through selective filtering of the \wmap\ maps.
However, since this analysis uses per-pixel fitting and unfiltered maps,
simulations are necessary to estimate any bias in $\beta^{pol}$ resulting from the presence of
loss-imbalance modes.  We use the simulations to derive a spatial template corresponding
to loss-imbalance related bias in $\beta^{pol}$, and to determine the significance of any correlation between this template and the data.

The estimated maximum amplitude of loss-imbalance 
modes at 23 and 33~GHz is lower than the instrument noise, but could possibly produce biases in the recovered
$\beta^{pol}$ of up to $\Delta\beta \sim 0.05$.
The 23 and 33~GHz mode morphologies are quite similar between the two bands, and attempting to individually estimate
contributions at each frequency with the data here is not possible. However, the frequency with the largest
potential uncertainty contribution is 23~GHz, because the sky signal is larger at that frequency compared to 33~GHz.  

We generate 1000 baseline realizations
of 1.4, 2.3, 23 and 33~GHz Q and U maps with instrument noise, CMB and synchrotron signal as estimated from 
the \planck\ FFP10 simulations (see Section~\ref{sec:hisig}).
Then a complementary set of 1000 `loss imbalance' simulations are generated, consisting of the 
same noise and signal components as the baseline set, plus a contribution\footnote{K-band template1 from \url{https://lambda.gsfc.nasa.gov/product/map/dr5/loss_imbal_template_r4_get.cfm}} 
to the simulated 23~GHz Q and U maps from loss imbalance.

\begin{figure}[ht]
    \centering
    \includegraphics[width=3.5in]{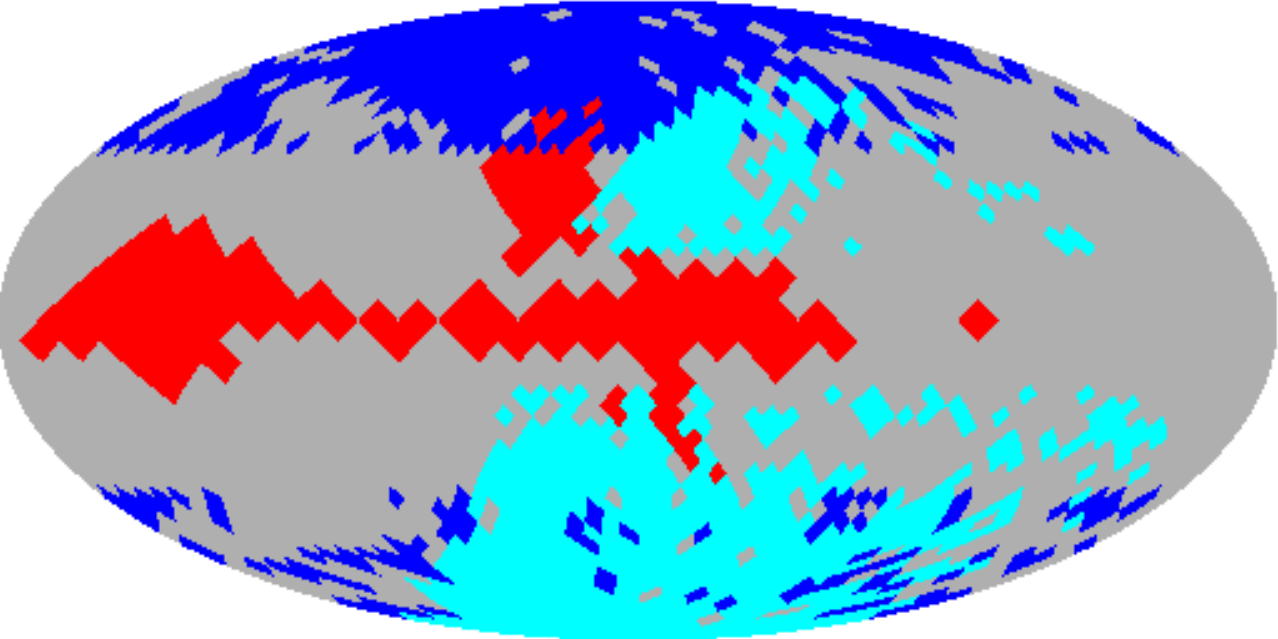}     
    \par
    \vspace{7mm}
    \includegraphics[width=3.5in]{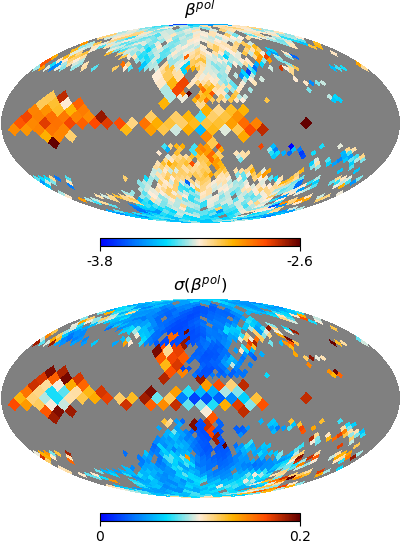}  
   \caption{
   {\it{Top:}} Color-coded map showing where the three separate analyses presented in Sections 3 and 4 contribute to the final
    composite $\beta^{pol}$ map. Pixels from the high SNR region analysis using \wmap\ 23 and 33 GHz bands are shown in red  
    Those from the higher latitude [1.4, 23, 33] are in blue, and pixels from the [2.3,23,33] region are in cyan.  Regions in gray are not included in the final map.
   {\it{Middle and Bottom:}} Composite map of polarized synchrotron spectral index (middle) and associated uncertainty (bottom) based on 
   analyses of 1.4, 2.3, 23 and 33 GHz maps. Pixels in gray are not analyzed. Section 5 describes the process by which the composite maps are created. 
    }
    \label{fig:bigger_merged_betal}
\end{figure}

Fits to the [1.4,23,33] and [2.3,23,33] frequency combinations are performed as described in Section~4.2
for the data, separately for the 1000 baseline simulations that do not include the loss imbalance signature, and for the 
1000 simulations that include the effect.  The spectral index map recovered from the simulations that include the loss imbalance term has a large-scale
systematic difference from that recovered from the baseline simulations.  From the mean of the difference between $\beta^{pol}$ maps recovered from these two simulation sets, we derive a map of the expected morphology that would be
introduced in $\beta^{pol}$ if a loss imbalance signature were present.  Recovered $\beta^{pol}$ maps are then fit with
the linear model $AT + c$, where $A$ scales the template $T$,
and $c$ is a constant.  

For the baseline simulations, the
expected value is $A=0.0$.  For the simulations including loss imbalance, we expect to recover $A=1$. However, this expectation 
only holds if the spectral index is constant over the whole sky, and the linear model is a good description of the data.
In the case of a spatially varying $\beta^{pol}$, the recovered value of $A$ will be biased because of chance correlations
with the spectral index morphology, and the partial sky coverage.  For simulations in which the input $\beta^{pol}$ is the same as that we derive for
the data (see Section~5), we recover $A = 1.43 \pm 0.21$ for the loss imbalance set, and $A = 0.47 \pm 0.21$ for the
null baseline set.  When we employ the same fitting procedure to the data, we obtain $A = 0.25 \pm 0.07$.  

The relatively low
value of $A$ obtained for the data implies a low contribution from the \wmap\ loss imbalance uncertainty $\Delta\beta \lesssim 0.01$, or alternatively that the fitting template obtained from simulations does not adequately match the data signature.  In
either case, we do not have sufficient evidence of a systematic bias in $\beta^{pol}$ resulting from loss imbalance
uncertainty, and do not include it in the estimated uncertainties. Residual quantification of the level of loss-imbalance signatures in \wmap\ bands will benefit from acquisition of additional independent data.

As described in Section~4.2, we have performed the [1.4,23,33] and [2.3,23,33] model fitting on a per-pixel basis, and 
ignored pixel-pixel covariances in the uncertainty estimation.  Our simulations,  which include the pixel-pixel correlations, 
confirm that we are not underestimating fitting uncertainties or biasing results because of this.

\section{Composite Synchrotron \texorpdfstring{$\beta^{pol}$}{betapol} Map} \label{sec:composite}

We create a composite $\beta^{pol}$ map at $N_{side}=16$ resolution by selectively populating pixels using the analysis results in 
Sections~\ref{sec:hisig} and \ref{sec:offplane}.
The selection process follows a hierarchy:  (1) fill all available pixels from the [2.3, 23, 33] GHz analysis, (2) next fill
remaining unpopulated pixels from the [1.4, 23, 33] GHz analysis, and (3) fill any remaining unpopulated pixels from those
in the [23, 33] GHz analysis.  In the case of the [23, 33] analysis, which was performed at $N_{side}=8$, we replicated the
value of each of the lower resolution pixels to fill the four corresponding pixels at the one step higher pixel resolution.
A color-coded map showing which of the three analyses was used to populate each pixel is shown in the top panel of
Figure~\ref{fig:bigger_merged_betal}.  Total sky coverage is $\sim 44$\%, with $\sim$73\% coverage of pixels for which  $|b| > 45^\circ$.

The composite $\beta^{pol}$ map is shown in the middle panel of Figure~\ref{fig:bigger_merged_betal}, with the uncertainty map in the
bottom panel.
Pixels in the uncertainty map are populated in the same manner as for the $\beta^{pol}$ map.  Uncertainties
computed for the [1.4, 23, 33] and [2.3, 23, 33] analyses include an absolute gain uncertainty contribution added in quadrature with the
statistical uncertainties.

\begin{figure}[ht]
    \includegraphics[width=3.5in,trim={1cm 0 0 0}, clip]{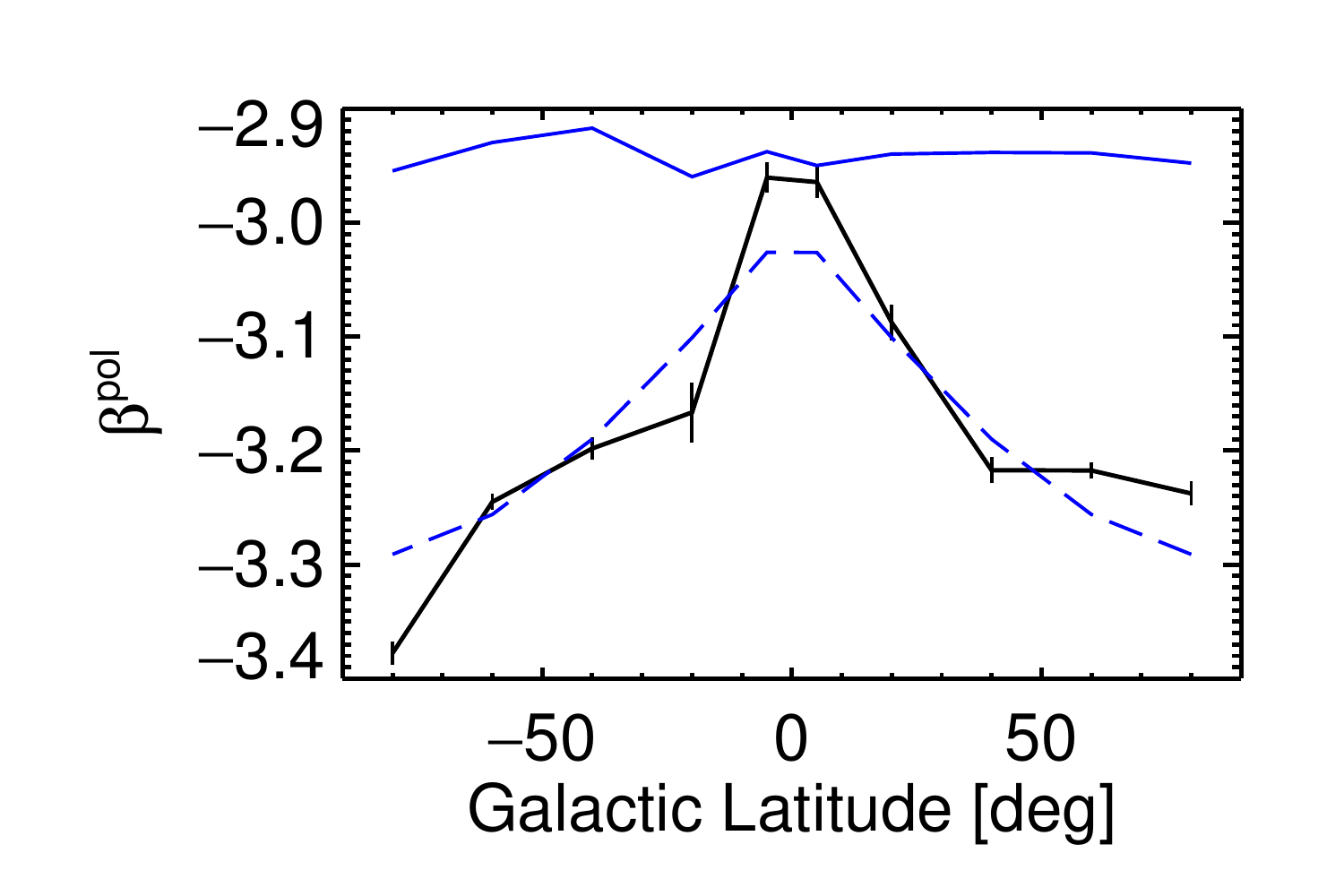}   
    \includegraphics[width=3.5in,trim={1cm 0 0 1cm}, clip]{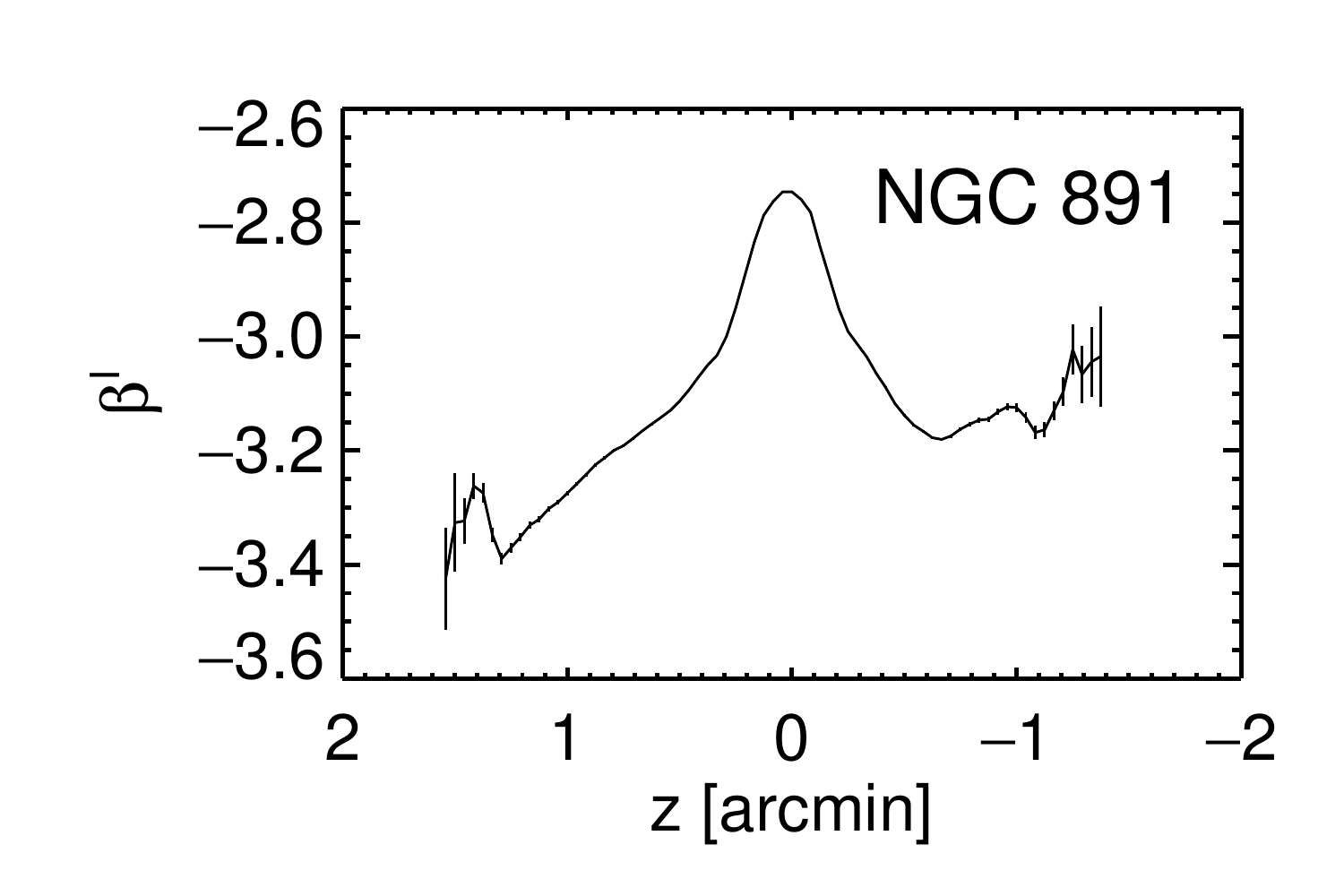}
    \caption{Top: Galactic latitude dependence of $\beta^{pol}$ as derived from binning 
    spectral index maps.  
    The profile in black (Table~\ref{tab:beta_profile_tab}) corresponds 
    to that derived from the composite spectral index map of Figure~\ref{fig:bigger_merged_betal}. 
    Blue traces correspond to binned profiles we generate from
    PySM \citep{thorne/etal:2017} synchrotron model 1 (solid) and model 2 (dashed).
    Bottom: Synchrotron intensity spectral index of Milky Way analog NGC 891 as a function of perpendicular distance z
    above and below mid-plane of that galaxy; one arcminute corresponds to $\sim$2.6 kpc.
    The profile is formed from the 1.5 and 6 GHz VLA non-thermal spectral index map of \citet{irwin/etal:2019}.  Unlike the models shown in blue in the top plot, the black data profiles in the top and bottom panels exhibit both latitudinal steepening and asymmetry.
    }
    \label{fig:beta_profile}
\end{figure}

The mean $\beta^{pol}$ latitudinal profile derived from the composite map is given in Table~\ref{tab:beta_profile_tab} and
plotted in the top panel of Figure~\ref{fig:beta_profile}.  Broad Galactic latitude bins ($20^\circ$ wide) are used in view of the
partial sky coverage.  Uncertainties are computed from the standard deviation within each bin, rather than using a weighted
average.  Because the $\beta^{pol}$ map lacks complete sky coverage, the latitude profile, particularly at mid-latitudes, is likely biased.  We suspect that any bias is toward larger values of $\beta^{pol}$ since coverage is weighted toward inner Galaxy regions.
The profile in Figure~\ref{fig:beta_profile} shows the same index steepening at $b= -80^\circ$ seen by \citet{fuskeland/etal:2019},
which is expected since S-PASS and 23~GHz data dominate the results for this region in both cases.
For contrast, latitudinal profiles for two synchrotron spectral index models from PySM (Python Sky Model, \citealt{thorne/etal:2017}) are shown in
blue.  The solid blue line is Model 1, which has been used for example in \planck\ FFP10 model simulations.  The dashed blue line
is for Model 2, a symmetric profile based on \wmap\ polarized foreground findings.  Neither model captures the detail seen
in the black trace of the data.

For further context, in the bottom panel of Figure~\ref{fig:beta_profile} we show the latitudinal profile of the synchrotron
intensity spectral index of the edge-on galaxy NGC~891, often cited as a Milky Way analog
\citep{hummel/etal:1991}.  We compute the NGC 891 profile from the 1.5 and 6 GHz non-thermal spectral index map of
\citet{irwin/etal:2019}, binning as a function of perpendicular distance z above and below the mid-plane of that galaxy.
Bins are in increments of 0.1 arminute, where one arcminute corresponds to $\sim$2.6 kpc (as a guide, an approximate scale-height for
cosmic-ray electrons in the Milky way is 1 kpc \citep{page/etal:2007}).  Although the NGC~891 profile is for intensity and not
polarization, there are similarities in high-latitude spectral index behavior, including similar spectral index values and a
north/south asymmetry.

\begin{table}[htb]
\centering
\caption{$\beta^{pol}$ Latitudinal Profile}
\label{tab:beta_profile_tab}

\begin{tabular}{c c }
\hline
Bin Center &  $\beta^{pol}$ \\ [0.5ex]
\hline\hline
   $80^\circ$  &  $-3.238 \pm 0.011$ \\
   $60^\circ$  &  $-3.217 \pm 0.007$  \\
   $40^\circ$  &  $-3.217 \pm 0.012$  \\
   $20^\circ$  &  $-3.088 \pm 0.016$ \\
   $5^\circ$   &  $-2.964 \pm 0.014$ \\
   $-5^\circ$  &  $-2.960 \pm 0.013$ \\
   $-20^\circ$  &  $-3.166 \pm 0.026$ \\
   $-40^\circ$  &  $-3.198 \pm 0.010$ \\
   $-60^\circ$  &  $-3.245 \pm 0.007$ \\
   $-80^\circ$  &  $-3.378 \pm 0.010$ \\

 \hline  
\end{tabular}

\end{table}

\section{Connecting Polarization and Intensity Data} \label{sec:intensity}

It is reasonable to consider augmenting the gaps in the composite $\beta^{pol}$ map with $\beta^I$ values derived
using temperature (intensity) maps of Galactic synchrotron 
emission\footnote{Note that a frequency-dependent polarization fraction can invalidate this assumption.}.
Full-sky temperature observations such as those from \wmap\ and
\planck\ LFI include contributions from multiple components in addition to synchcrotron, including CMB, free-free, thermal dust
and spinning dust emission.
Component separation studies limited to these frequencies are affected by degeneracies between multiple component spectral energy
distributions.
At frequencies $\lesssim 5$ GHz, foreground emission is much stronger than the CMB, and thermal and spinning dust emission are
expected to be subdominant to synchrotron and free-free \citep{harper/etal:2022}, which mitigates the component separation challenge.  
Unfortunately, calibration of ground-based observations at these lower frequencies is a difficult enterprise, and currently available
data limit effective use of these frequencies for diffuse synchrotron spectral index determination, despite a number of efforts
in the literature.

To illustrate the problem, we construct a synchrotron intensity model from a low-frequency template and extrapolate that template
to other frequencies assuming that the $\beta^{pol}$ we derived above is applicable to temperature observations.  We then compare
that model to publicly available temperature maps at 1.4 and 2.3~GHz.  As with previous investigations
(e.g., \citealt{bennett/etal:2003c, bennett/etal:2013, planck/10:2015, planck/04:2018}), we choose the Haslam 408 MHz map
\citep{haslam/etal:1982} as the
synchrotron template.  For consistency with the
\citet{planck/10:2015} component separation, we use the destriped version of \citet{remazeilles/etal:2015}, from which we remove an 8.9~K extragalactic background offset.  We also remove an estimate of free-free emission using the model of \citet{planck/10:2015}, but our
results are not substantially affected by that choice.  Single masked pixels in the $\beta^{pol}$ map have been inpainted with the mean 
from neighboring pixels in this simulation,
for the purpose of allowing a more contiguous visual exposition of data $-$ model residuals.

\begin{figure}[t]
    \centering
    \includegraphics[width=3.5in]{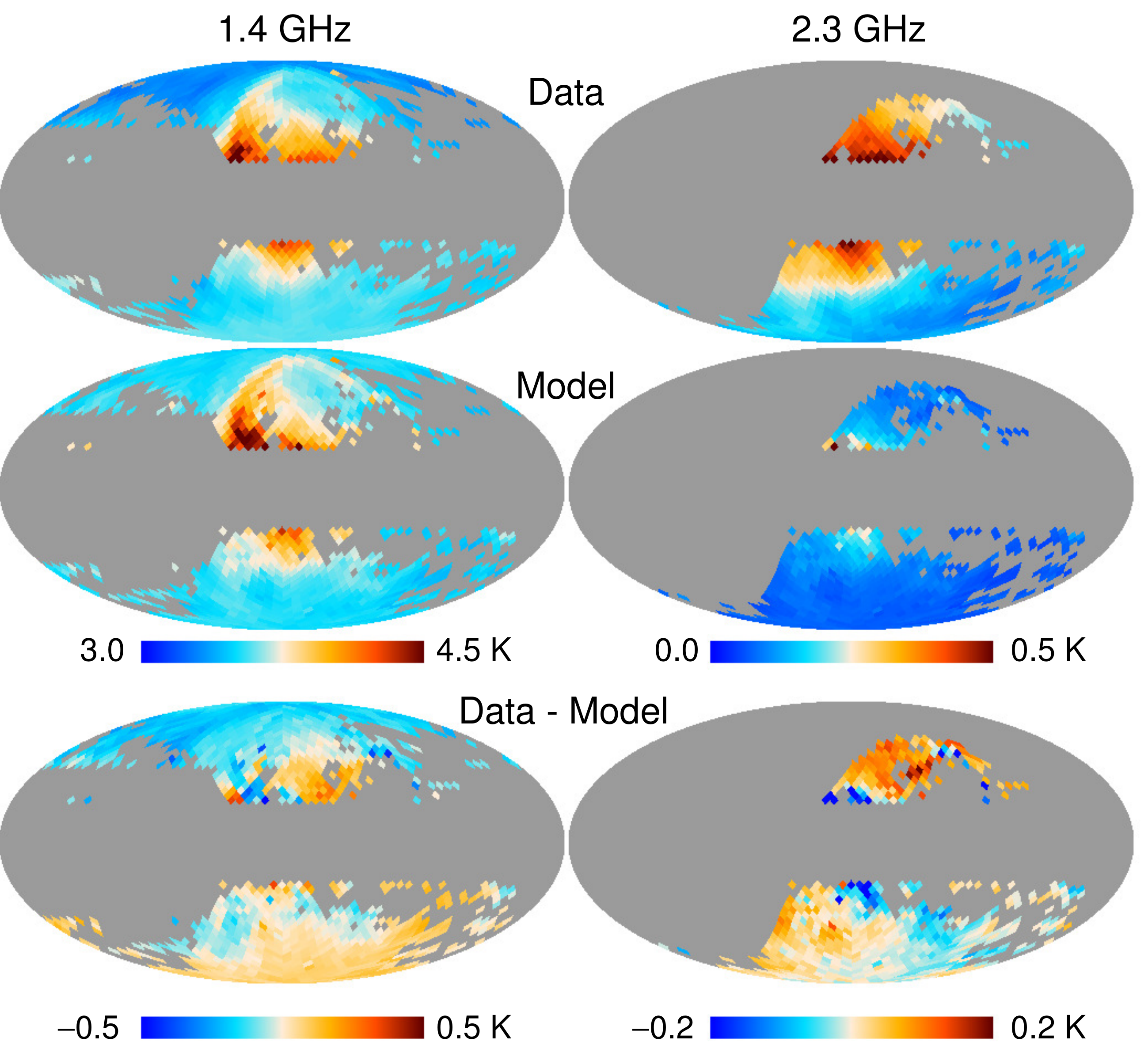}
    \caption{Observed temperature maps at 1.4 GHz (top left) and 2.3 GHz (top right) are modeled as a summation of synchrotron and free-free components in the middle panel, assuming a 408 MHz synchrotron template and a $\nu^{\beta^{pol}}$ frequency dependence.  An extragalactic background component
    is included in the 1.4 GHz model, but not for 2.3 GHz (see text).  The bottom panel is the difference between the
    data and the model.  No scaling was applied to the 1.4 GHz model, but the 2.3 GHz model is multiplied by a factor of 3
    in order to produce the residual in the bottom right.
    Calibration discrepancies are an issue affecting attempts
    use these low-frequency intensity datasets to compare $\beta^I$ with $\beta^{pol}$.
    }
    \label{fig:intensity_models}
\end{figure}

The top left panel in Figure~\ref{fig:intensity_models} shows the 1.4~GHz intensity map from the combined northern and southern
surveys of \citet{reich/reich:1986} and \citet{reich/testori/reich:2001}, available from CADE. 
The quoted absolute gain uncertainty is roughly 5\%, with a zero-point uncertainty of 500 mK.
The middle left panel shows the corresponding model sky emission consisting of a synchrotron component computed from $T_{synch}^{408} (1.41/0.408)^{\beta^{pol}}$, a free-free component estimated using the model of \citet{planck/10:2015}, and
an added monopole of 3300 mK, which accounts for the extragalactic background and zero-point uncertainties.  The two panels have
visual similarities, but it is clear that there is a high latitude north-south asymmetry in the observations compared to the model.  The difference
between the observations and model are shown in the bottom left panel.  The magnitude of the high latitude differences in the north would require a
difference of order $0.5$ between $\beta^I$ and $\beta^{pol}$, such that a spectral index map produced from
0.408 and 1.4 GHz would deviate significantly compared to that seen in $\beta^{pol}$.  We note that the original northern sky
determination of $\beta^I$ using two these frequencies 
\citep{reichbeta:1988} indicated a typical $\beta^I \sim -2.5$ at $b \sim 40^\circ$, as opposed to $\beta^{pol}$ near $-3.2$.

The right half of Figure~\ref{fig:intensity_models} compares the more recent southern sky 2.3~GHz intensity map from the S-PASS survey \citep{carretti/etal:2019} against a model prediction. 
The 2.3~GHz model includes synchrotron and free-free, as did
the 1.4~GHz model, but not an extragalactic component, since
the monopole of the S-PASS map is calibrated to Galactic emission levels with an uncertainty of 70 mK.  There is a significant
discrepancy in brightness between the model and observations, at a level greater than the 5-10\% absolute gain uncertainty quoted for
the S-PASS and Haslam surveys.
A linear correlation between the
model and the observations indicates that the 2.3 GHz observations are $\sim 2-3$ times brighter than the model; we do not give an exact number 
because the correlation is neither tight nor strictly linear.  The linear correlation derives an offset near 50~mK,
which is consistent with the zero-point uncertainty.  We adjust the model based on the correlation slope and offset, and subtract it from the observations, with the result shown in the bottom right panel.
This residual has a large-scale spatial pattern, but one quite different from that seen at 1.4 GHz.
The scaling factor is unexpected, but as we can reproduce the polarization fraction values in Figure 27 of \citet{carretti/etal:2019}, there does not seem to be an error in our use of the delivered data files.
The high scaling factor implies a calibration inconsistency between the older (408 MHz, 1.4 GHz) intensity data, and the 
newer 2.3~GHz intensity survey.  The recent independent comparison of $\beta^I$ derived using either Haslam 408~MHz or
preliminary C-BASS\footnote{C-Band All Sky Survey} 5~GHz data as the synchrotron template
\citep{harper/etal:2022} would seem to indicate that the S-PASS intensity data are in discord with the other surveys.  There
is no indication that S-PASS polarization data are in substantial disagreement on large scales with 1.4~GHz
polarization data, however 
(Section~\ref{sec:offplane_radio}).

In short, new polarization and intensity data from ongoing experiments such as C-BASS \citep{c-bass:2018J} and 
QUIJOTE\footnote{QUI JOint TEnerife} \citep{quijote:2015}, as well as upcoming experiments, promise to provide valuable new constraints on the CMB foregrounds.

\section{Foreground Removal Implications} \label{sec:implications}

Current data limitations prevent a well-constrained determination of polarized synchroton foreground contributions 
over the entire sky.  As noted by \citet{bicep/keck:2021} and \citet{fuskeland/etal:2019}, contributions at larger
spatial scales ($\ell < 20$) of interest to e.g. LiteBIRD\footnote{(Lite) B-mode polarization \& Inflation from cosmic background Radiation Detection }\citep{litebird:2020} and CLASS\footnote{Cosmology Large Angular Scale Surveyor} \citep{class:2022}
are of most concern, where the foreground power is greatest.  
The primary issue of cosmological importance is the accuracy to which foregrounds can be characterized and removed 
in relation to the CMB signal.
In that respect, the spatial scale and range of $\beta^{pol}$ variations is a key factor.  
A significant large-scale $\beta^{pol}$ gradient with Galactic latitude is demonstrated in Figure~\ref{fig:beta_profile}.  
Longitudinal variations are more subtle, but present.
Two examples of selected regions taken from the composite $\beta^{pol}$ map follow.

\begin{figure}
    \centering
    \includegraphics[width=3.25in]{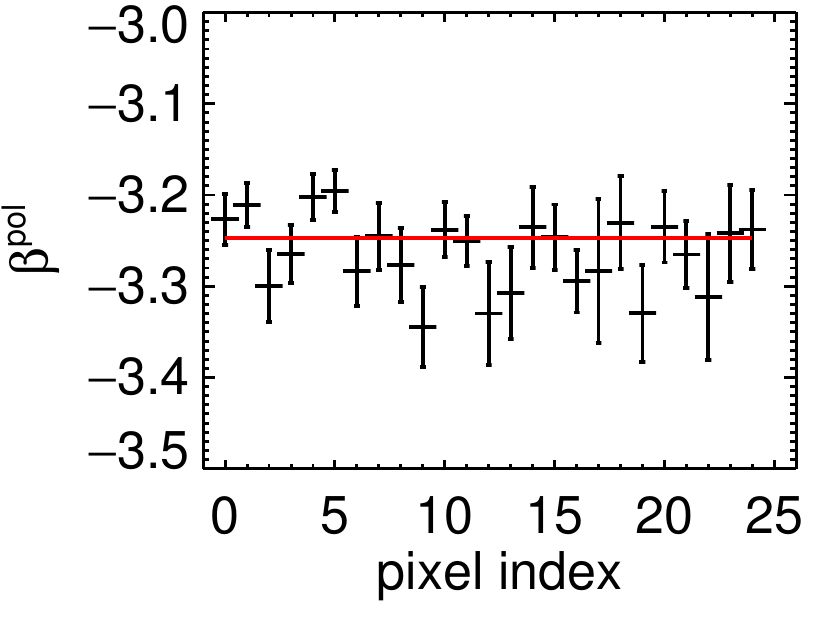}
   \caption{Plot of $\beta^{pol}$ for each of the 25 unmasked $N_{side}=16$ pixels within the BICEP2 footprint.   Pixels
   are plotted in arbitrary order and given index numbers 0 to 24.  The weighted mean ($-3.25$) is shown as the horizontal red line.  There is no
   statistical evidence for spectral index variation within the region: this is consistent with the assumed lack of
   variation used in the BICEP2/Keck analysis \citep{bicep/keck:2021}.
   }
    \label{fig:bicep}
\end{figure}

In the first example, we investigated the extent to which we could detect spectral variations in the BICEP2\footnote{Background Imaging of Cosmic Extragalactic Polarization 2} \citep{bicep2/2014} field,  based on the 25 unmasked $N_{side}=16$ pixels
that lie within the survey footprint (a $400$ deg$^2$ patch centered near $l \sim 316^\circ, b \sim -58^\circ$, as taken from the
LAMBDA\footnote{Legacy Archive for Microwave Background Data Analysis}
Footprint Tool\footnote{\url{https:/lambda.gsfc.nasa.gov/toolbox/footprint/}}).  We plot the $\beta^{pol}$ values and uncertainties for these pixels in
Figure~\ref{fig:bicep}.  Uncertainties increase with decreasing Galactic longitude across the field, but the spectral index distribution is consistent
with no variation, with a probability to exceed (PTE) = 0.21.  The mean value we compute for this region is $-3.25 \pm 0.04$(statistical)$\pm 0.02$(systematic).
This determination is consistent with the $\beta^{pol}=-3.22 \pm 0.06$ value adopted by the BICEP Collaboration for the entire field
\citep{bicep/keck:2021}.
The determination is also consistent with the mean latitudinal variation in $\beta^{pol}$ 
predicted for this field from the  Figure~\ref{fig:beta_profile} data points. We compute the expected range in $\beta^{pol}$ 
based on the range of latitudes in the BICEP2 field and a linear interpolation of
the values in Table~\ref{tab:beta_profile_tab}. 
The computed peak-to-peak latitudinal variation in $\beta^{pol}$ is $\sim 0.05$,
which is not distinguishable from a constant given the data uncertainties.

\begin{figure}
    \centering
    \includegraphics[width=3.25in]{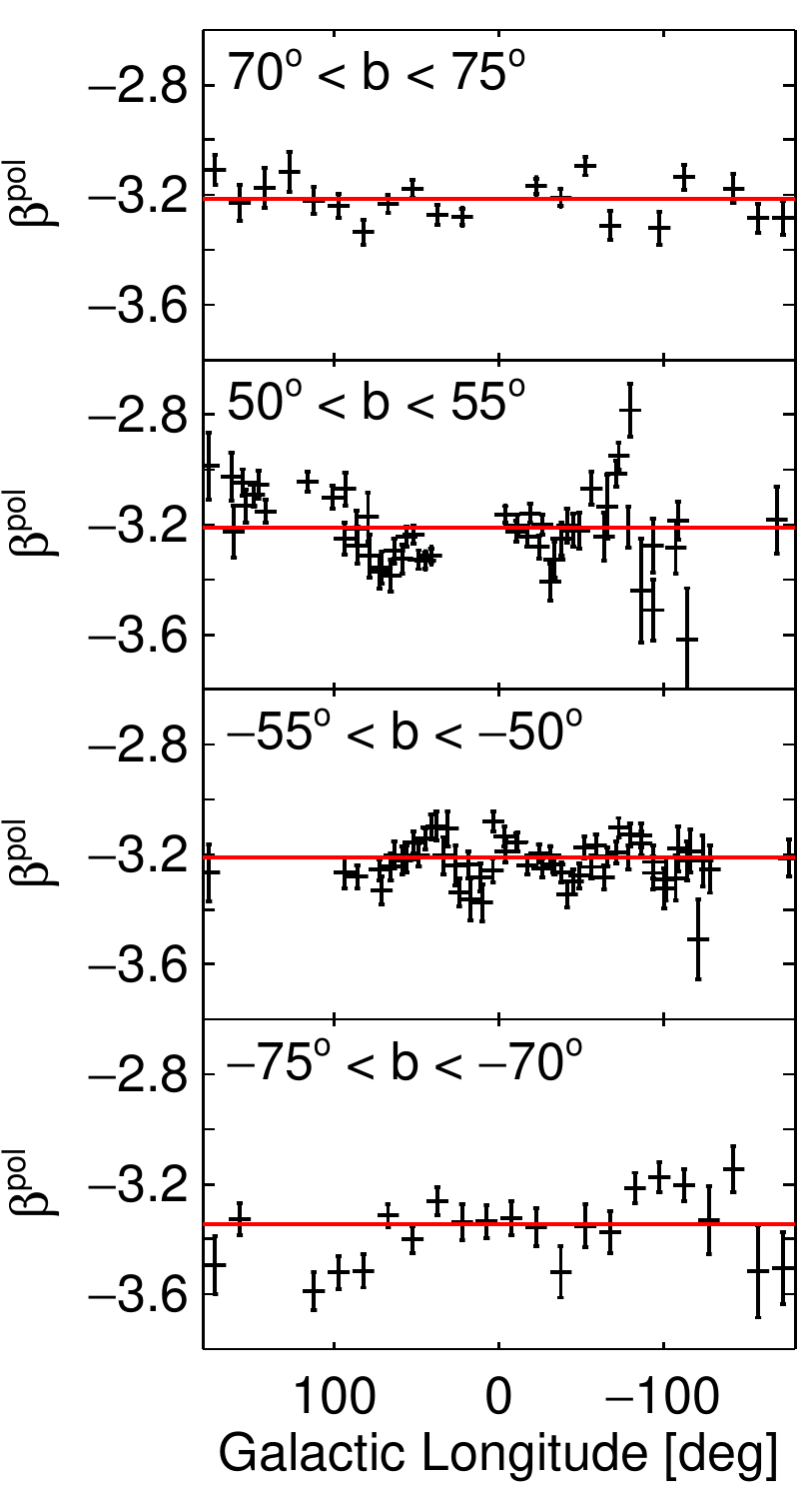}  
    \caption{Slices through the composite $\beta^{pol}$ map at four fixed Galactic latitudes: $b=\pm 52.5$ and $\pm 72.5$.  The latitude
    bins are $5^\circ$ wide, and each panel plots $\beta^{pol}$ within the slice as a function of Galactic longitude. The red horizontal line in each panel corresponds to the weighted mean of all unmasked pixels within the latitude bin.  PTE values for the bottom three panels
    are all $< 10^{-7}$ and strongly reject being modeled as a constant value.  The PTE for the top panel is $2.5\times10^{-4}$ and also
    inconsistent with a constant within errors.
    }
    \label{fig:longitude_slices}
\end{figure}

The second example illustrates longitudinal spectral index behavior at four fixed Galactic latitudes. We choose 
high latitudes for which the composite $\beta^{pol}$ map has a reasonably large range of longitude coverage. These fixed latitude
slices, shown in Figure~\ref{fig:longitude_slices}, are centered on $b = \pm 52.5^\circ$ and $b = \pm 72.5^\circ$, and include
all unmasked $\beta^{pol}$ map pixels within each of the $5^\circ$ wide latitude bins.  For each of the four slices, we test the null
hypothesis that spectral indices at all longitudes are consistent with a constant value given by their weighted mean (indicated by the red horizontal
line in the Figure).  The PTE for the top panel is $2.5\times10^{-4}$, and those for the remaining three panels are all of orders of magnitude lower and
thus show strong statistical evidence for longitudinal variation.

\begin{figure}
    \centering
    \includegraphics[width=3.25in]{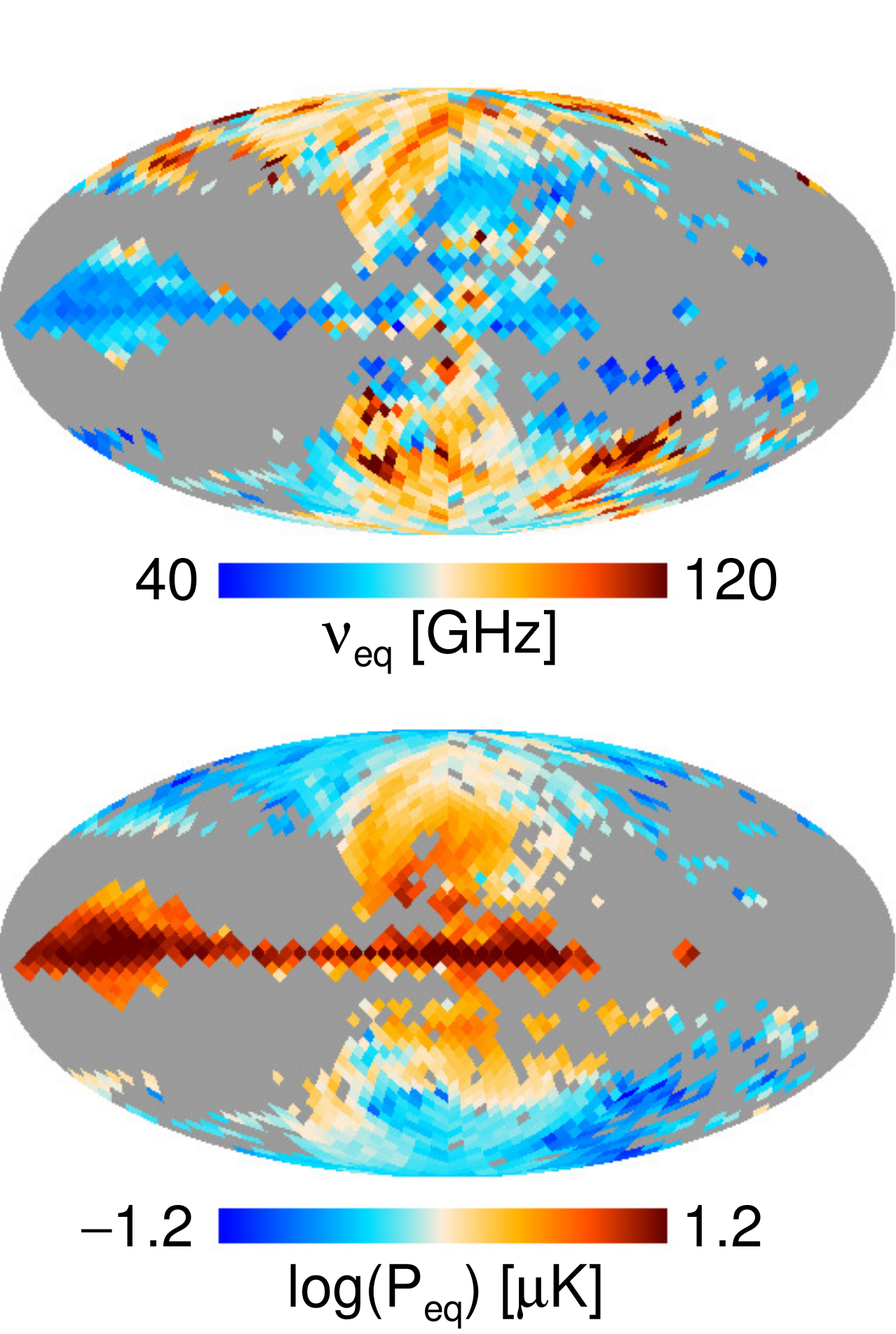}    
    \caption{{\it{Top: }}Map of the frequency $\nu_{eq}$ at which synchrotron and thermal dust emission contribute equally to
    the total polarized foreground, assuming the synchrotron $\beta^{pol}$ from Figure~\ref{fig:bigger_merged_betal} and a thermal dust spectral index of 1.55
    \citep{planck/04:2018}.  
    {\it{Bottom: }}The sum of synchrotron and dust polarized intensity computed at $\nu_{eq}$, in thermodynamic units, for the same spectral index choices
    described for the top row.  The scale is logarithmic, with all blue shades indicating $P_{eq} < 1 \mu$K.
    }
    \label{fig:crossover}
\end{figure}

The role that these variations play in synchrotron foreground removal depends upon the range of frequencies being analyzed, the selected sky region, the
specific removal technique being employed, and the potential removal accuracy possible in context of instrumental noise.  
The scale of spatial variations shown in this paper favor techniques adaptable to variations of order 0.05 - 0.1  on scales
of several degrees, and tend to  disfavor removal methods that rely on an assumed constant spectral energy distribution over large areas.
For example,
a simulated noiseless case that extrapolates a 5~GHz template to determine the 90~GHz synchrotron component over $|b| > 50^\circ$, but uses a fixed spectral index
of $-3.2$, would produce a median residual value corresponding to a tensor-to-scalar ratio $r = 0.05$, with some residuals exceeding
$r = 0.5$.  With the recent BICEP2/Keck upper limit $r \le  0.036$ \citep{bicep/keck:2021}, such residuals
conflict with the goal of detecting B-modes at 90~GHz.  However, as we saw for the smaller BICEP2 footprint, the use of a constant $\beta^{pol}$ was within the uncertainties derived in this paper.  

Not all experiments have the option to survey the entire sky and must choose cosmologically interesting fields.  In the top panel of Figure~\ref{fig:crossover},  we
have computed the frequency at which polarized synchrotron and thermal dust contribute equally to the total foreground emission, for those regions of the sky
that we have analyzed. The synchrotron model is based on the \wmap\ 23~GHz polarized intensity and the map of $\beta^{pol}$ shown in Figure~\ref{fig:bigger_merged_betal}.
The thermal dust polarized contribution is modeled using a modified blackbody with spectral index 1.55 and dust temperature 19.6~K \citep{planck/04:2018}, and with 
the amplitude corresponding to the NPIPE 353~GHz full-mission polarized intensity.  For off-plane regions, we find $\nu_{eq} = 79 \pm 13$~GHz.
A similar mean value was found by \citet{planck/04:2018} based on rms amplitudes computed over sky fractions from 0.27 to 0.83.
In the Figure~\ref{fig:crossover} top panel, pixels with
$\nu_{eq} \gtrsim 90$ tend to have lower polarized dust emission.  The bottom panel of the Figure shows the total foreground polarized
intensity
at $\nu_{eq}$ in thermodynamic temperature units.  Most, but not all, higher latitude regions have lower foreground contributions.
While we have estimates for uncertainties in $\beta^{pol}$, uncertainties in the polarized dust spectral index are not well characterized
\citep{osumi/etal:2021}, and thus these maps only serve as a guide rather than a complete picture.

\section{Conclusions}  \label{sec:concl}

Well constrained characterization of the amplitude and spectral energy distribution of polarized synchrotron emission is 
a necessity for future
high sensitivity cosmological experiments observing frequencies $< 150$ GHz.  
This paper explores the extent to which current publicly available datasets
with significant sky coverage can advance that goal.
The analysis reaches three main conclusions:
\begin{enumerate}
  \item The polarized synchrotron spectral index, $\beta^{pol}$, is not a constant over
    the entire sky, and the variations we derive are not well matched by frequently used models 
    such as those in PySM.
  \item Current public data are insufficient to characterize $\beta^{pol}$ for the future, for three reasons: a lack of full sky coverage; conflicting results between experiments; and a lack of sensitivity to support future B-mode experiments.
  \item Some sky regions are well enough measured to provide some guidance for some experiments, such as in the BICEP2/Keck analysis \citep{bicep/keck:2021}.
\end{enumerate}
We summarize our findings in greater detail below.

The public data we analyze are comprised of Stokes Q and U maps from the
1.4~GHz surveys of DRAO and Villa Elisa \citep{wolleben/etal:2010, testori/etal:2008},
the S-PASS 2.3~GHz southern sky survey \citep{carretti/etal:2019}, the \WMAP\ 23 and 33~GHz 9-year maps \citep{bennett/etal:2013},
and, in specific cases, \Planck\ 30~GHz maps from Public Release 3 (PR3, \citealt{planck/01:2018}), Public Release 4 (PR4, aka NPIPE, \citealt{npipe:2020}, and
BeyondPlanck \citep{beyondplanck_pipeline:2020}.

A significant part of our analysis involves data selection and consistency tests between independent datasets that have differing
sensitivities and calibration uncertainties.  The presence of Faraday depolarization in both the 1.4~GHz and 2.3~GHz maps requires
that we divide the analysis into two spatial domains: the high SNR regions of the Galactic plane and spurs (Section~\ref{sec:hisig}),
and lower SNR off-plane regions (Section~\ref{sec:offplane}).  Within these two regimes, we found additional calibration discrepancies
that required further selection choices:

\begin{itemize}[leftmargin=0.0in]
    \item  In the high SNR Galactic plane and spurs  (Figures~\ref{fig:beta_kka_k30} and \ref{fig:beta_regions}),
    we found significantly different results for $\beta^{pol}$ in the plane when using pairs of frequencies from \wmap\ only 
    (23 and 33~GHz) and from \wmap\ 23 ~GHz in combination with any of the \planck\ 30~GHz processing pipeline versions.  
    The discrepancy traces to large-scale systematic differences between the \wmap\ 23~GHz and LFI 30~GHz maps, which have
    been shown \citep{npipe:2020} to change as a function of the mapmaking algorithm used to process the 30~GHz data.
    For this reason, we sacrifice the SNR advantages that would be gained from a combination of \WMAP\ and \planck\ LFI
    data, and our results are based on the \wmap\ [23, 33]~GHz maps only.
    
    \item  In the lower SNR off-plane domain, \wmap\ and \planck\ data lack the requisite SNR to determine $\beta^{pol}$ on
    all but very large sky patches.  It is therefore desirable to include lower-frequency radio data (in our case, 1.4 and 2.3~GHz)
    where the synchrotron SNR is much higher.  In Section~\ref{sec:offplane_radio}, we show that systematic calibration differences
    exist between the 1.4 and 2.3~GHz surveys for $b < -70^\circ$ (Figure~\ref{fig:southpole_im}),
    and argue against using these two frequencies alone for a southern sky determination of $\beta^{pol}$.
    
    \item  In light of the above two points, our off-plane analysis fits a parametrized sky model to two three-frequency map combinations:
    [1.4, 23, 33]~GHz and [2.3, 23, 33]~GHz (Section~\ref{sec:offplane_3freq}).  Although we fit the entire sky, not all pixel fits 
    have the same quality.  We find some pixels for which there is strong disagreement between [1.4, 23, 33] and [2.3, 23, 33].  We further downselect pixels based on depolarization, systematic measurement errors, and/or low SNR regions.

\end{itemize}

Following our data quality assessment, we construct a composite map of the polarized synchrotron spectral index $\beta^{pol}$
(Section~\ref{sec:composite}).
The composite map is populated following a hierarchy that first fills all available pixels from the [2.3, 23, 33] GHz analysis,
next fills remaining unpopulated pixels from the [1.4, 23, 33] GHz analysis, and finally fills any remaining unpopulated
pixels from those in the [23, 33] GHz analysis (top panel of Figure~\ref{fig:bigger_merged_betal}).
This results in $\beta^{pol}$ coverage over 44\% of the sky (73\% for $|b| > 45^\circ$),
with a pixel resolution of $\sim 3.7^\circ$ ($\sim 7.3^\circ$ in the plane).  The maps and associated uncertainties are shown in
Figure~\ref{fig:bigger_merged_betal}.  Uncertainties include statistical noise and instrument absolute gain uncertainties. 
We searched for potential bias in the composite $\beta^{pol}$ map arising from \wmap\ mirror transmission efficiency differences,
but did not definitively detect any (Section~\ref{sec:sim_imbal}).

Variation in $\beta^{pol}$ is an important factor in synchrotron foreground removal.  Because of data limitations, we are unable
to discern spectral curvature, and have assumed a pure power law frequency dependence in constructing the composite
spectral index map.  Based on the composite map, we characterize spectral spatial variations in both Galactic latitude and longitude:

\begin{itemize}[leftmargin=0.02in]
    \item In the Galactic plane and spurs, we find $ -3.2 < \beta^{pol} \lesssim -3$ for much of the region, but with a flatter value for
    the Fan Region in the outer Galaxy.
    
    \item We find a clear gradient in $\beta^{pol}$ with Galactic latitude, but the gradient is not symmetric between northern and southern hemispheres.  The mean $\beta^{pol}$ latitude profile indicates
    spectral index steepening with increasing latitude south of the Galactic plane with $\Delta \beta^{pol}=0.4$, and a smaller steepening of $0.25$ in the north. Near the south Galactic pole  the polarized synchrotron spectral index is $\beta^{pol} \approx -3.4$.
    As discussed in the text, indications of a gradient have previously been reported in the literature.  We note that the
    latitude profile we derive has the potential to bias high, particularly in the mid-latitude regions where spatial coverage is 
    predominantly from inner Galaxy regions.

    \item For those high latitude sky regions included in our composite map, we find longitudinal variations in $\beta^{pol}$ of order 0.05-0.10 about the mean latitudinal value. This result
    has a greater dependence on the accuracy of the data uncertainties used in the analysis, and leaves unanswered the applicability
    to fainter high latitude regions for which we could not sufficiently constrain $\beta^{pol}$.  

    \item We find $\beta^{pol}$ within the BICEP2/Keck survey footprint.to be consistent with a constant value, 
    $\beta^{pol} = -3.25 \pm 0.04$ (statistical) $\pm 0.02$ (systematic), in accord with the value adopted in
    \citet{bicep/keck:2021}.

\end{itemize}

Since data selection criteria did not allow a full-sky determination of $\beta^{pol}$, we assessed the possibility of filling coverage
gaps using a spectral index determined from  0.408, 1.4 and 2.3~GHz intensity data. At these frequencies,
the diffuse Galactic sky signal is dominated by synchrotron emission, with some contribution from free-free. Unfortunately, 
a preliminary comparison between these datasets only 
served to emphasize systematic differences between currently available observations at these frequencies (Section~\ref{sec:intensity}, Figure~\ref{fig:intensity_models}), 
and no $\beta^I$ determinations were included in our analysis.

At $N_{side}=16$ resolution, the $\beta^{pol}$ map is of most interest to ongoing and future experiments that
survey large sky areas and target low multipole CMB reionization and recombination signatures
($2 \lesssim \ell \lesssim 100$), such at LiteBIRD and CLASS.   For those portions of the sky
that it covers, the $\beta^{pol}$ map may be used directly, although 
resolution and sensitivity limitations restrict its applicability. For those regions lacking coverage, the map
serves to anticipate the level of synchrotron spectral variation that future instrument design
and sky cleaning algorithms must account for.  We express this roughly 
as a
latitudinal variation overlaid with longitudinal variations of order 0.05 - 0.1 on few degree scales.

Ultimately however,
the limitations and inconsistencies among datasets encountered in this work
make clear the value of additional independent surveys at multiple frequencies.  
These additional surveys are necessary to provide increased sensitivity at low SNR high latitude
locations and provide sufficient frequency coverage to assess e.g. spectral curvature.
The
frequency window of most utility for high latitudes is $ 2.3 < \nu < 30$ GHz, and especially between 10-20~GHz 
where depolarization is minimized while still ensuring synchrotron signal dominance.
There are ongoing
ground-based projects working to augment frequency coverage in this window. Calibration from the ground is substantially more difficult than in space however, and many analyses will
still rely on \wmap\ and \planck\ data as key frequencies. Although the absolute calibration for these two surveys is sub-percent, the SNR 
at some high-latitude locations is insufficient for precision determination of $\beta^{pol}$,
and additional high-quality data are needed.  

We plan to make the composite $\beta^{pol}$ map available through LAMBDA\footnote{\url{https://lambda.gsfc.nasa.gov}} upon publication.

\vspace*{0.15in}
This research was supported in part by NASA grants NNX17AF34G, 80NSSC19K0526, 80NSSC20K0445, and
80NSSC21K0638.
This research has made use of NASA's Astrophysics Data System Bibliographic Services. 
Some of the results in this paper have been derived using the \textsl{healpy} and \textsl{HEALPix} package.
We acknowledge the use of the 
Legacy Archive for Microwave Background Data Analysis (LAMBDA), part of the High Energy Astrophysics Science Archive Center (HEASARC). 
HEASARC/LAMBDA is a service of the Astrophysics Science Division at the NASA Goddard Space Flight Center.  
We also acknowledge use of the \textit{Planck} Legacy Archive. \textit{Planck} is an ESA science mission with instruments and contributions 
directly funded by ESA Member States, NASA, and Canada.
This work has made use of S-band Polarisation All Sky Survey (S-PASS) data.

\end{document}